\documentclass[prd,aps,twocolumn,a4paper,floatfix,showpacs,nofootinbib]{revtex4-1}

\usepackage{graphicx,psfrag}
\usepackage{mathrsfs}
\usepackage{amsmath,amsfonts,amssymb,pifont}
\usepackage{multirow,enumerate}
\usepackage{comment,hyperref}
\usepackage{color}
\usepackage{acronym}
\usepackage{xspace}
\usepackage[normalem]{ulem}

\newacro{BH}{black hole}
\newacro{NS}{neutron star}
\newacro{PN}{Post-Newtonian}
\newacro{BBH}{binary black hole}
\newacro{BNS}{binary neutron star}
\newacro{EOB}{effective-one-body}
\newacro{NR}{numerical relativity}
\newacro{GW}{gravitational wave}
\newacro{EOS}{equation-of-state}

\newcommand{\be}{\begin{equation}}
\newcommand{\ee}{\end{equation}}
\newcommand{\bea}{\begin{eqnarray}}
\newcommand{\eea}{\end{eqnarray}}
\newcommand{\bel}{\begin{align}}
\newcommand{\eel}{\end{align}}

\def\GMc2{{\rm G M_{\odot} c^{-2}}}

\def\Qw{Q_{\hat{\omega}}}
\def\hw{{\hat{\omega}}}

\def\mo{\hat{\omega}}
\def\ct{\tilde{c}}

\def\nt{\tilde{n}}
\def\dt{\tilde{d}}
\def\NRtidal{\texttt{NRTidal}\xspace}
\def\SEOBNR{\texttt{SEOBNRv4}\xspace}
\def\SEOBNRROM{\texttt{SEOBNRv4\_ROM}\xspace}
\def\SEOBNRROMNRtidal{\texttt{SEOBNRv4\_ROM\_NRTidal}\xspace}
\def\PhenomDNRtidal{\texttt{PhenomD\_NRTidal}\xspace}
\def\PhenomPNRtidal{\texttt{PhenomPv2\_NRTidal}\xspace}
\def\PhenomP{\texttt{PhenomPv2}\xspace}
\def\PhenomD{\texttt{PhenomD}\xspace}
\def\TEOBResumS{\texttt{TEOBResumS}\xspace}
\def\SEOBNRv4T{\texttt{SEOBNRv4T}\xspace}
\def\bam{{\textsc{BAM}}\xspace}
\def\TaylorF{\texttt{TaylorF2}\xspace}
\def\TaylorT{\texttt{TaylorT4}\xspace}
\def\TaylorFT{\texttt{TaylorF2$_{\rm Tides}$}\xspace}
\def\TaylorTT{\texttt{TaylorT4$_{\rm Tides}$}\xspace}

\usepackage{color}

\definecolor{cyan}{rgb}{0,0.9,0.9}
\definecolor{orange}{rgb}{0.9,0.5,0}
\definecolor{magenta}{rgb}{1,0,1}
\definecolor{purple}{rgb}{0.8,0.4,0.8}
\definecolor{gray}{rgb}{0.5,0.5,0.5}
\definecolor{mygreen}{rgb}{0.1,0.8,0.1}
\definecolor{darkblue}{rgb}{0.0,0.0,0.6}

\newcommand{\AEIHannover}{Max Planck  Institute for Gravitational Physics
(Albert Einstein Institute), Callinstr.~38, 30167 Hannover, Germany}
\newcommand{\UniHannover}{Leibniz Universit\"at Hannover, D-30167 Hannover, Germany}

\begin{document}

\title{Matter imprints in waveform models for neutron star binaries: tidal and self-spin effects} 

%main work done by
\author{Tim \surname{Dietrich}$^{1,2}$}
\author{Sebastian \surname{Khan}$^{3,4}$}
\author{Reetika Dudi$^{5}$}
\author{Shasvath J. Kapadia$^6$}
\author{Prayush Kumar$^7$}
\author{Alessandro \surname{Nagar}$^{8,9,10}$}
\author{Frank Ohme$^{3,4}$}
\author{Francesco Pannarale$^{11}$}
\author{Anuradha Samajdar$^1$}
%supporters
\author{Sebastiano \surname{Bernuzzi}$^{12,13}$}
\author{Gregorio \surname{Carullo}$^{14}$}
\author{Walter \surname{Del Pozzo}$^{14}$}
\author{Maria Haney$^{15}$}
\author{Charalampos Markakis$^{16,17}$}
\author{Michael \surname{P\"urrer}$^{2}$}
\author{Gunnar \surname{Riemenschneider$^{9,18}$}}
\author{Yoshinta Eka Setyawati$^{3,4}$}
\author{Ka Wa \surname{Tsang}$^{1}$}
\author{Chris Van Den Broeck$^{1,19}$}

\affiliation{${}^1$ Nikhef, Science Park, 1098 XG Amsterdam, The Netherlands}
\affiliation{${}^2$ Max Planck Institute for Gravitational Physics (Albert Einstein Institute),
                    Am M\"uhlenberg 1, Potsdam 14476, Germany}
\affiliation{${}^3$ \AEIHannover}
\affiliation{${}^4$ \UniHannover}
\affiliation{${}^5$ Theoretical Physics Institute, University of Jena, 07743 Jena, Germany}
\affiliation{${}^6$ Center for Gravitation, Cosmology, and Astrophysics, University of Wisconsin-Milwaukee, Milwaukee, Wisconsin 53201, USA}
\affiliation{${}^7$ Cornell Center for Astrophysics and Planetary Science, Cornell University, Ithaca, New York 14853, USA}
\affiliation{${}^8$ Centro Fermi - Museo Storico della Fisica e Centro Studi e Ricerche “Enrico Fermi”, Rome, Italy}
\affiliation{${}^{9}$ INFN Sezione di Torino, Via P. Giuria 1, 10125 Torino, Italy}%
\affiliation{${}^{10}$ Institut des Hautes Etudes Scientifiques, 91440 Bures-sur-Yvette, France}%
\affiliation{${}^{11}$ School of Physics and Astronomy, Cardiff University, The Parade, Cardiff CF24 3AA, UK}
\affiliation{${}^{12}$ Dipartimento di Scienze Matematiche Fisiche ed Informatiche, Universit\'a di Parma, I-43124 Parma, Italia}
\affiliation{${}^{13}$ Istituto Nazionale di Fisica Nucleare, Sezione Milano Bicocca, gruppo collegato di Parma, I-43124 Parma, Italia}
\affiliation{${}^{14}$ Dipartimento di Fisica “Enrico Fermi”, Università di Pisa, Pisa I-56127, Italy}
\affiliation{${}^{15}$ Physik-Institut, Universit\"at Z\"urich, Winterthurerstrasse 190, 8057 Z\"urich, Switzerland}
\affiliation{${}^{16}$ NCSA, University of Illinois at Urbana-Champaign, 1205 W Clark St, Urbana, Ilinois 61801, USA}
\affiliation{${}^{17}$ DAMTP, Centre for Mathematical Sciences, Wilberforce Road, University of Cambridge, Cambridge, CB3 0WA, UK}
\affiliation{${}^{18}$ Dipartimento di Fisica, Universit`a di Torino, via P. Giuria 1, I-10125 Torino, Italy}
\affiliation{${}^{19}$ Van Swinderen Institute for Particle Physics and Gravity,
University of Groningen, Nijenborgh 4, 9747 AG Groningen, The Netherlands}%

\date{\today}

\begin{abstract}
The combined observation of gravitational and electromagnetic
waves from the coalescence of two neutron stars marks the beginning
of multi-messenger astronomy with gravitational waves (GWs).
The development of accurate gravitational waveform models is a crucial
prerequisite to extract information about the properties of
the binary system that generated a detected GW signal. In
binary neutron star systems (BNS), tidal effects also need
to be incorporated in the modeling for an accurate waveform
representation. Building on previous work [Phys.~Rev.~D96~121501],
we explore the performance of inspiral-merger waveform models
that are obtained by adding a numerical relativity
(NR) based approximant for the tidal part of the phasing
(NRTidal) to existing models for nonprecessing and
precessing binary black hole systems (SEOBNRv4, PhenomD and
PhenomPv2), as implemented in the LSC Algorithm Library Suite.
The resulting BNS waveforms are compared and contrasted
to target waveforms hybridizing
NR waveforms, covering the last $\sim 10$ orbits up to merger and
extending through the postmerger phase, with inspiral waveforms
calculated from 30Hz obtained with TEOBResumS. The latter is
a state-of-the-art effective-one-body waveform model that blends
together tidal and spin effects. 
We probe that the combination of the
PN-based self-spin terms and of the NRTidal description is 
necessary to obtain minimal mismatches ($\lesssim 0.01$)
and phase differences ($\lesssim 1$~rad) with respect to the
target waveforms. However, we also discuss possible improvements and 
drawbacks of the NRTidal approximant in its current form, 
since we find that it tends to overestimate the tidal interaction 
with respect to the TEOBResumS model 
during the inspiral. 
\end{abstract}

\maketitle

\section{Introduction}

On August 17, 2017, the \ac{GW} detector network
formed by the Advanced LIGO and Virgo interferometers detected
GW170817, the first \ac{GW} signal consistent with the inspiral and merger of a
\ac{BNS} system~\cite{TheLIGOScientific:2017qsa}.
In addition to the \ac{GW} signal, astronomers observed the short
gamma-ray burst GRB 170817A~\cite{Monitor:2017mdv} and the transient AT 2017gfo
in the X-ray, ultraviolet, optical,
infrared, and radio bands~\cite{GBM:2017lvd,2017Sci...358.1556C} from the same source.
This joint detection initiated a new era of
multi-messenger astronomy.
From this single observation
it was already possible to prove
that \ac{BNS} mergers are central engines for short gamma-ray bursts,
and that they produce heavy elements which give rise to
electromagnetic counterparts known as
kilonovae or macronovae.
Additionally, measurements of the speed of GWs~\cite{Monitor:2017mdv} as
well as of the Hubble constant were performed~\cite{Abbott:2017xzu}.  Finally,
new constraints on the
unknown \ac{EOS} of cold
matter at supranuclear densities were determined,
e.g.~\cite{TheLIGOScientific:2017qsa,Radice:2017lry,
Bauswein:2017vtn}.
Due to the increasing sensitivity of advanced \ac{GW} detectors
over the next years, multiple detections of merging
BNSs are expected in the near future~\cite{Abbott:2016ymx}.

Extracting information about the properties of the binary system
from \ac{GW} detector data is crucial for the field of \ac{GW} astronomy.
Source properties are generally inferred via a coherent, Bayesian
analysis that involves repeated cross-correlation of the measured \ac{GW}
strain with predicted waveforms~\cite{Veitch:2014wba}.  Therefore, the
computation of individual waveforms needs to be efficient and fast.
Furthermore, in contrast to \ac{BBH} systems, which are
usually detectable for the last few orbits before merger, \ac{BNS} systems
are visible by \ac{GW} detectors for several seconds or even minutes before
the merger.  Consequently, computational efficiency is even
more important for \ac{BNS} waveform approximants than for \ac{BBH} systems.
On the other hand, the computed waveforms need to be an accurate representation of the
binary system to allow for correct estimates of the source properties,
such as the masses and spins, and, in the case of \ac{BNS} systems, the
internal structure of the stars.

Significant progress in modeling \ac{BNS} systems was accomplished over the last few years, capturing the strong-gravity and
tidally dominated regime of the late-inspiral.
State-of-the-art tidal waveform models in the time domain have been
developed in \cite{Bernuzzi:2014owa, Hinderer:2016eia, Steinhoff:2016rfi} and are based on the
\ac{EOB} description of the general-relativistic
two-body problem \cite{Buonanno:1998gg, Damour:2009wj}.
While this approach is powerful and accurately describes the waveform
up to the moment of merger for a variety of binary configurations
within the uncertainty of state-of-the-art \ac{NR} simulations,
there are \ac{BNS} parameter space regions for which recent numerical simulations
suggest that further improvements of the tidal \ac{EOB} models are
necessary~\cite{Hotokezaka:2015xka, Dietrich:2017feu}.
But, the biggest disadvantage of this approach is the high computational cost
required to compute a single waveform. While applying reduced-order-modeling
techniques~\cite{Lackey:2016krb} allows to overcome this issue,
it also adds additional complexity.
Therefore, modeling techniques complementary
to \ac{EOB}, e.g.~\cite{Lackey:2013axa, Pannarale2013a, Barkett:2015wia, Lange:2017wki}, are needed, 
especially because \ac{PN} approximants
become increasingly inaccurate towards the merger,
e.g.~\cite{Bernuzzi:2012ci,Favata:2013rwa,Wade:2014vqa,Hotokezaka:2016bzh}.

In Ref.~\cite{Dietrich:2017aum}, the authors propose the first closed-form
tidal approximant combining \ac{PN}, tidal \ac{EOB}, and \ac{NR} information.
This waveform model was implemented in the LSC Algorithm Library
(LAL) Suite~\cite{lalsuite} to support the analysis of
GW170817~\cite{TheLIGOScientific:2017qsa}.  Specifically, a particular
version of this tidal approximant was added to the point-mass dynamics
described by the spin-aligned \ac{EOB} model of~\cite{PhysRevD.95.044028}
and by the phenomenological, frequency-domain
approach of~\cite{PhysRevD.93.044006, PhysRevD.93.044007,PhysRevLett.113.151101}.
Very recently~\cite{Kawaguchi:2018gvj} also developed
a tidal approximant in the frequency domain combining
\ac{EOB} and \ac{NR} information, for a comparison between
the~\NRtidal model and the model of~\cite{Kawaguchi:2018gvj}
we refer the reader to Appendix~E of~\cite{Kawaguchi:2018gvj}.

In addition to the tidal interaction, another EOS-dependent effect
that distinguishes BNSs from BBH binaries is the deformation that
the star acquires due to its own rotation (self-spin or monopole-quadrupole
terms), that eventually leaves an imprint on the gravitational-wave
signal~\cite{PhysRevD.57.5287}.
The outcome of the leading-order (LO) PN-based description of this
effect on \ac{EOS} measurements has been investigated in recent
works~\cite{Agathos:2015uaa,Harry:2018hke} and it has been 
incorporated, in resummed form, in the \TEOBResumS EOB
model~\cite{Nagar:TEOBResumS} and in the SEOBNRv4T~\cite{Hinderer:2016eia,Steinhoff:2016rfi} model. 

The main goal of this paper is to asses the quality of the implementation
of tidal effects described by the \NRtidal model in the LALSuite as
well as of the PN-description of EOS-dependent self-spin effects.
This is done by comparing LALSuite BNS waveforms to hybrid waveforms
obtained by matching together NR waveforms,  covering the last
$\sim 10$ orbits up to merger and extending through the postmerger
phase, with inspiral waveforms (calculated from 30Hz) obtained
with \TEOBResumS.
In Sec.~\ref{sec:models}, we discuss the tidal phase correction which
is the key element of transforming a \ac{BBH} baseline model to obtain
\ac{BNS} waveforms, as well as the \ac{BBH} baseline models we employed.
We discuss the \ac{NR} simulations and hybrids used for our tests
in Sec.~\ref{sec:nrhybrids}, and validate the models in the frequency
domain and the time domain in Sec.~\ref{sec:FDchecks} and
in Sec.~\ref{sec:TDchecks}, respectively, using either mismatches or
direct phase comparisons. Finally, Sec.~\ref{sec:nrtides_systematics}
points out systematic effects that are present in the current
implementation of \NRtidal that may affect parameter estimation studies
introducing biases. 
Our conclusions are collected in Sec.~\ref{sec:conclusion}.

Throughout this article geometric units are used by setting $G=c=M_\odot=1$.

\section{IMR\_NRTidal models}
\label{sec:models}

\subsection{Model description}

\subsubsection{Numerical relativity and effective-one-body tuned tidal phase correction}

In contrast to the \ac{BBH} case, waveforms describing the
emission from BNS systems need to include tidal effects that
incorporate the fact that each star gets tidally
polarized due to the tidal field of the
companion~\cite{Damour:1983LesHouches,Hinderer:2007mb,Damour:2009vw,Binnington:2009bb}.
In the following, we include tidal effects by means of the method outlined
in~\cite{Dietrich:2017aum}, where the tidal phase has been extracted from
the tidal \ac{EOB} model of~\cite{Bernuzzi:2014owa} and high-resolution \ac{BNS} \ac{NR} waveforms.
The procedure is outlined here; the interested reader can find a detailed discussion in Ref.~\cite{Dietrich:2017aum}.

We consider a binary with total mass $M=M_A + M_B$, with the convention
that $M_A\geq M_B$. Defining the complex GW as $h(t)=A(t)e^{-{\rm i}\phi(t)}$,
the time-domain phase $\phi(t)$ is assumed to be given by the following
PN-inspired sum of individual contributions
\begin{equation}\label{eq:phi_omg}
 \phi(\mo) \approx \phi_{\rm pp} (\mo) + \phi_{\rm spin}(\mo) + \phi_{\rm tides}(\mo) \ ,
\end{equation}
where $\mo = M\omega=M\partial_t\phi(t)$ is the dimensionless \ac{GW} frequency,
$\phi_{\rm pp} (\mo)$ denotes the nonspinning, point-particle,
contribution to the overall phase, $\phi_{\rm spin}$ corresponds
to contributions caused by spin effects, and $\phi_{\rm tides}$
corresponds to contributions caused by tidal effects.
As shown in~\cite{Dietrich:2017aum,Dietrich:2018upm} one finds that
during the last orbits before merger, i.e., the regime accessible by \ac{NR} simulations,
and for dimensionless spin magnitudes up to $S_{A,B}/M_{A,B}^2\sim 0.1$
current state-of-the-art \ac{NR} simulations are
not capable of revealing tides-spin coupling,
which supports the specific form of Eq.~\eqref{eq:phi_omg} and
the absence of a $\phi_{\rm spin\leftrightarrow tides}$ term.

Non-spinning tidal contributions enter the phasing at the 5PN
order~\footnote{As mentioned in the introduction, \ac{EOS} dependent
phase corrections depending on the self-spin interaction, the
quadrupole-monopole terms, appear already at 2PN~\cite{Poisson:1997ha}
(see also Ref.~\cite{Levi:2014sba} for the NLO contributions).
As shown in Sec.~\ref{sec:FDchecks} and discussed, e.g.~in~\cite{Harry:2018hke},
those effects are important for spinning \ac{BNS} systems even with dimensionless
spins $\sim 0.1$.}. The fully known next-to-leading order (NLO)
PN expression of the tidal contribution
(TaylorT2 approximant)~\cite{Damour:2009wj,Vines:2010ca,Vines:2011ud,Wade:2014vqa} is
\begin{equation}\label{eq:T2}
\phi_{\rm tides} = - \kappa_{\rm eff}^T \frac{c_{\rm Newt}
x^{5/2}}{X_A X_B}  (1 + c_1 x )  \ ,
\end{equation}
with $x(\mo)=(\mo/2)^{2/3}$, $c_{\rm Newt} = -13/8$ and $X_{A,B}=M_{A,B}/M$.
The NLO tidal correction to the phasing, $c_1$,
is for the equal-mass case $c_1 = 1817/364$.

The parameter $\kappa_{\rm eff}^T$ characterized tidal effects and reads
\begin{equation}\label{eq:kappa}
\kappa^T_{\rm eff} = \frac{2}{13} \left[
\left(1+12\frac{X_B}{X_A}\right)\left(\frac{X_A}{C_A}\right)^5 k^A_2 +
 (A \leftrightarrow B)
\right]  \ ,
\end{equation}
where $C_{A,B}\equiv M_{A,B}/R_{A,B}$ are the compactnesses of the stars at isolation,
and $k^{A,B}_2$ the Love numbers describing the static quadrupolar deformation
of one body in the gravitoelectric field of the companion~\cite{Damour:2009vw}.
The tidal parameter $\kappa^T_{\rm eff}$ is connected to $\tilde{\Lambda}$ used to characterize
tidal effects in Ref.~\cite{TheLIGOScientific:2017qsa} via $\tilde{\Lambda}=16/3\kappa^T_{\rm eff}$,
once one has defined the tidal polarizability parameters as $\Lambda_{A,B}=2/3 k_2^{A,B}/C_{A,B}^5$.

\begin{figure}[t]
\includegraphics[width=0.5\textwidth]{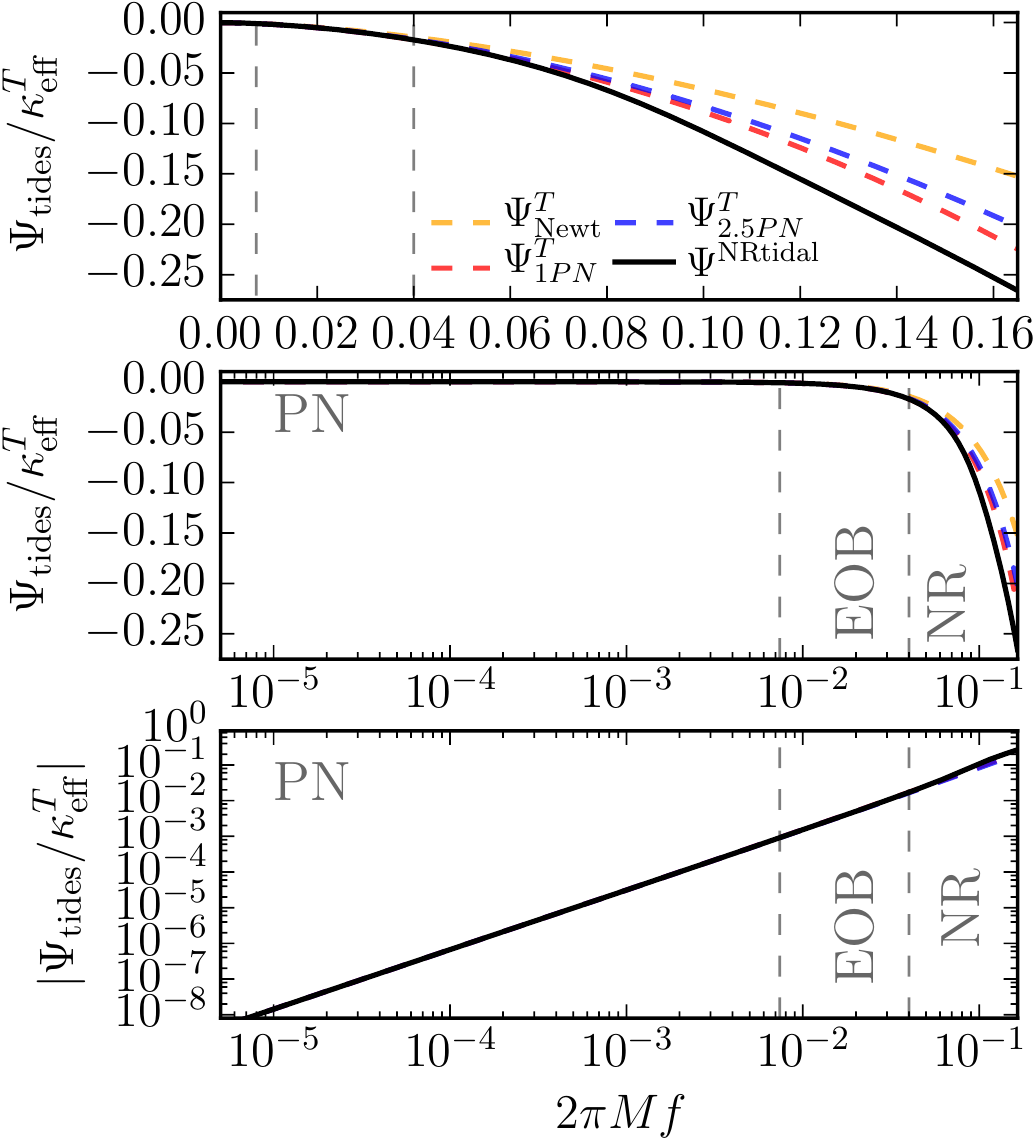}
\caption{Tidal phase correction for the \NRtidal (black) and various \ac{PN} models 
truncated at the corresponding PN orders from Eq.~\eqref{eq:psiT_exact}.
From top to bottom, the phase correction is shown on a linear,
on a semi-logarithmic,
and on a double logarithmic scale, respectively.
The vertical, dashed lines mark regions in which we calibrate the \NRtidal
model to \ac{PN}, \ac{EOB}, and \ac{NR} waveforms. }
\label{fig:tidal_phase}
\end{figure}

An effective representation of the tidal effects coming beyond
the NLO Eq.~\eqref{eq:T2} can be obtained using
the following expression
\begin{equation}\label{eq:tD_general}
\phi_{\rm tides} = - \kappa_{\rm eff}^T
\frac{c_{\rm Newt} x^{5/2}}{X_A X_B} P^{\rm NRTidal}_\phi(\mo) \ ,
\end{equation}
where $P^{\rm NRTidal}_\phi(\mo)$ is fitted to \ac{PN}, \ac{EOB}, and \ac{NR} waveforms
in such a way that for $\mo \leq 0.0074$ Eq.~\eqref{eq:T2}
is used to determine $P_\phi(\mo)$,
for $\mo \in[0.0074,0.04]$ the tidal \ac{EOB} waveforms
of~\cite{Bernuzzi:2014owa,Damour:2014sva,Nagar:2015xqa}
are used, and, finally, Richardson-extrapolated
\ac{NR} data~\footnote{For calibration of the \NRtidal model to NR dat 
the configurations 
SLy$_{1.35|1.35}^{0.00|0.00}$, 
H4$_{1.37|1.37}^{0.00|0.00}$, 
and MS1b$_{1.35|1.35}^{0.00|0.00}$ have been used, cf.~\cite{Dietrich:2017aum} 
for a detailed discussion.} of~\cite{Dietrich:2017aum} are used for
$\mo \in[0.04,0.17]$, cf.~Fig.~\ref{fig:tidal_phase}.
The final expression for $P_\phi(\mo)$ is represented
by a rational function of the form
\begin{small}
\begin{equation}
\label{eq:fitT}
P^{\rm NRTidal}_\phi(\mo) = \frac{1  + n_1 x + n_{3/2} x^{3/2} +
n_2 x^{2} + n_{5/2} x^{5/2}+n_3 x^3}{1+ d_1 x + d_{3/2} x^{3/2}}.
\end{equation}
\end{small}
We require that Eq.~\eqref{eq:fitT} reproduces Eq.~\eqref{eq:T2} at low
frequencies which is ensured when $d_1=(n_1-c_1)$.
The coefficients are given by
$(n_1,n_{3/2},n_2,n_{5/2},n_3) = (-17.941,57.983,-298.876,964.192,-936.844)$,
and $d_{3/2}=43.446$.

The tidal phase correction in the frequency domain is computed from
the time-domain approximant via stationary phase
approximation~\cite{Damour:2012yf}. The integration is performed
numerically and the numerical data are represented as
\begin{align}
\label{eq:NRTidal}
\Psi^{\rm NRTidal}(f) & = -\kappa_{\rm eff}^T\frac{\ct_{\rm Newt}}{X_A X_B} x^{5/2}P^{\rm NRTidal}_\Psi ,
\end{align}
with
\begin{align}
\label{eq:Pnrtides}
P^{\rm NRTidal}_\Psi=\frac{1  + \nt_1 x + \nt_{3/2} x^{3/2} + \nt_2 x^{2} + \nt_{5/2} x^{5/2}}{1+ \dt_1 x + \dt_{3/2} x^{3/2}},
\end{align}
where $x=x(f)$, $\ct_{\rm Newt} = 39/16$ and $\dt_1 = \nt_1 - 3115 /1248$,
while the remaining parameters are determined by fitting. They read
$(\nt_1,\nt_{3/2},\nt_{2},\nt_{5/2})=(-17.428,31.867,-26.414,62.362)$
and $\dt_{3/2}=36.089$. Similarly to Eq.~\eqref{eq:fitT},
$\dt_1=(\nt_1-\ct_1)$ ensures that the NLO tidal term is
correctly recovered.

Equation \eqref{eq:NRTidal} gives then the final \NRtidal
correction which can be added to any tidal-free waveform model.
We will discuss all current point-mass baseline models to
which the  \NRtidal correction has been added in the next subsection.
A comparison between $\Psi^{\rm NRTidal}$, Eq.~\eqref{eq:NRTidal},
and \ac{PN} tidal predictions is shown in Fig.~\ref{fig:tidal_phase}.
We show $\Psi_{\rm tides}/\kappa_{\rm eff}^T$
on a linear, semi-logarithmic, and double logarithmic scale.
Additionally, we mark the intervals for which we used \ac{PN}, \ac{EOB}, \ac{NR} datapoints
to tune the \NRtidal model. As seen in the bottom panel of
Fig.~\ref{fig:tidal_phase}, only the last part of the inspiral is
affected by the calibration to \ac{EOB} and \ac{NR} waveforms.

In this respect, we want to stress that the choice of explicitly incorporating
analytical NLO tidal information in the above formulas was made mainly for simplicity 
and to reduce the number of parameters in Eq.~\eqref{eq:fitT}.
In fact, tidal information beyond NLO are available~\cite{Damour:2012yf},
see the discussion in Sec.~\ref{sec:tidesbeyondNLO}.

\subsubsection{Point-particle baseline models}

Models for the \ac{GW} signal from \ac{BBH} systems are under active development
for use in the analysis of Advanced LIGO-Virgo data.
Typically, the complex \ac{GW} strain $h$ is decomposed into a spin-weight -2
spherical harmonic basis, i.e.,
\begin{equation}
h(t; \theta, \phi) = \sum\limits_{\ell \geqslant 2} \sum\limits_{m = -\ell}^{\ell} h_{\ell, m}(t) Y^{-2}_{\ell,m}(\theta, \phi)\,,
\label{eq:hlm}
\end{equation}
where for comparable mass, non-precessing systems the $\ell=|m|=2$ multipoles
are dominant.

Two of the models used in the analysis of Advanced LIGO-Virgo data
are the \ac{EOB} model \SEOBNR \cite{PhysRevD.95.044028} and
the phenomenological model \PhenomD \cite{PhysRevD.93.044006, PhysRevD.93.044007}.
These are aligned-spin models for the  $\ell=|m|=2$ multipoles that
use \ac{PN}/\ac{EOB} to describe the early inspiral and then calibrate
model coefficients to \ac{NR} waveforms to predict the late-inspiral, merger
and ringdown.

The agreement between these aligned-spin \ac{BBH} models was quantified in
\cite{PhysRevD.95.044028}. With increasing mass ratio and for positively aligned spins with a
magnitude above $\sim 0.5$, their agreement drops and their mismatch exceeds $3\%$. However,
the two \ac{BBH} models agree in large regions of the BNS parameter space, and we therefore
expect negligible differences between
the models when we compare them against the set of \ac{NR} waveforms we use
in this study.

Contrary to the aligned-spin waveform models,
\PhenomP includes precession effects.
It is built upon the assumption that the spin-orbit coupling can be
approximately separated into components
parallel and perpendicular to the instantaneous
orbital angular momentum, with the former influencing the rate of inspiral
and the latter driving the precessional
motion~\cite{Kidder:1995zr,Apostolatos:1994mx,Buonanno:2002fy,Schmidt:2010it,Schmidt:2012rh,Schmidt:2014iyl}.

\subsection{LALSuite implementation}

\subsubsection{Addition of tidal phase}

The simplicity of the tidal correction given via Eq.~\eqref{eq:NRTidal}
allows us to add $\Psi^{\rm NRTidal}$ to any frequency domain waveform model
which accurately represents the point-particle or \ac{BBH} coalescence.
We construct tidal models of the \SEOBNR and \PhenomD models called
\SEOBNRROMNRtidal and \PhenomDNRtidal, respectively.

The construction permits a particularly simple implementation
where we add the $\Psi^{\rm NRTidal}$ correction along with a
suitable amplitude $A^{\rm NRTidal}$ to the point-particle \ac{GW} polarizations
obtained from the {\tt LALSimulation} library according to the following
equation
\begin{equation}
h^{\rm tidal} = h^{\rm pp} \cdot \left( A^{\rm NRTidal} e^{-i \Psi^{\rm NRTidal}} \right),
\label{eq:corr}
\end{equation}
where $A^{\rm NRTidal}$ is a function that smoothly turns off the
waveform shortly after the termination frequency described below.

To construct a precessing tidal waveform approximant from the \PhenomP BBH baseline model,
we add the tidal correction to the underlying spin-aligned \PhenomD model on which \PhenomP
is built on \emph{before} we rotate the waveform
according to the angles that describe the precession dynamics~\cite{Schmidt:2012rh,Schmidt:2014iyl}.
At leading order, the tidal effects decouple from the precessional motion
and the resulting precessing waveform model should still be valid.
The inspiral part of \PhenomP (and \PhenomD) is based upon the \TaylorF approximant.
This allows us to include the NLO effect
due to the spin-induced quadrupole-monopole interaction \cite{PhysRevD.57.5287}, 
which currently is only included in the \PhenomPNRtidal
model\footnote{Note that this is caused by historical reasons.
  The NLO quadrupole-monopole term could have been added also 
  to the other \NRtidal waveform models}.
This modifies the spin contribution to
the quadrupole moment, which is a function of the \ac{EOS}.
Here we utilize the universal relations of \cite{YAGI20171} to relate the
tidal deformability parameters $\Lambda_{1,2}$ to the
spin-induced quadrupole-monopole terms. These terms in the \PhenomPNRtidal
waveform model occur at the NLO in the inspiral phase.

\subsubsection{Termination criterion for tidal correction}

The tidal phase correction in Eq.~\eqref{eq:NRTidal} only describes the
inspiral part of the \ac{BNS} coalescence.
Therefore, an additional criterion where to stop the computation of the waveform is required.
We relate this termination criterion to the merger frequency.
As outlined in~\cite{Bernuzzi:2014kca}, (for moderate mass ratios) the merger frequency of \ac{BNS}
systems is a function of
\begin{equation}
\kappa^T_{2} = 2 \left[
\frac{X_B}{X_A} \left(\frac{X_A}{C_A}\right)^5 k^A_2 +
\frac{X_A}{X_B} \left(\frac{X_B}{C_B}\right)^5 k^B_2 \right].
\label{eq:kappa2}
\end{equation}
Note that in the LAL implementation $\kappa_{\rm 2}^T$  is substituted by $\kappa_{\rm eff}^T$
for simplicity without introducing noticeable differences.
Based on a large set of \ac{NR} simulations,
the proposed fit of~\cite{Bernuzzi:2014kca}
was extended to include high-mass ratio
systems, e.g.~\cite{Dietrich:2015pxa,Dietrich:2016hky}, and reads
\begin{equation}
 \mo =
 \mo_0\sqrt{\frac{X_B}{X_A}}
 \frac{1+ n_1 \kappa_2^T + n_2 (\kappa_2^T)^2}
 {1+d_1 \kappa_2^T + d_2 (\kappa_2^T)^2}, \label{eq:Momega_mrg}
\end{equation}
with $n_1=3.354 \times 10^{-2}, n_2 = 4.315 \times 10^{-5},
d_1 = 7.542 \times 10^{-2}, d_2 = 2.236 \times 10^{-4}$.
The parameter $\mo_0=0.3586$ in Eq.~\eqref{eq:Momega_mrg} is chosen such
that for equal-mass cases $q=M^A/M^B=1$ and $\kappa_2^T \rightarrow 0$
the nonspinning \ac{BBH} limit is
recovered~\cite{Bernuzzi:2014kca,Healy:2017mvh}.

We use Eq.~\eqref{eq:Momega_mrg} to determine the end of the waveform and
taper the signal using a Planck taper \cite{McKechan:2010kp}.
The taper begins at the estimated merger frequency and ends at $1.2$ times the merger
frequency. Because of the smooth frequency evolution even after
the moment of merger~\cite{Dietrich:2017aum}, we do not expect to introduce non-negligible errors
due to evaluating Eq.~\eqref{eq:NRTidal} after the merger frequency.

\section{Target hybrid waveforms}
\label{sec:nrhybrids}

As mentioned in the introduction, to validate the various phasing models
built using the prescription given by Eq.~\eqref{eq:corr} we compare them
against complete BNS waveforms that are constructed stitching together
the analytical waveforms constructed within the EOB approach with
waveforms obtained through \ac{NR} simulations. In the following
three paragraphs we briefly discuss: (i) the properties of the analytical
\TEOBResumS EOB model; (ii) the properties of the NR waveforms; (iii) the
procedure to hybridize EOB to NR waveforms in the overlapping frequency
region ($\sim$ the last $10$ orbits before merger) so as to obtain complete
waveforms that cover the full frequency range, from the early, quasi-adiabatic
inspiral, up to the postmerger phase.

\begin{table*}[t]
  \centering
  \caption{\ac{BNS} configurations.
    The name of the configuration, following the notation
    EOS$_{M_A|M_B}^{\chi_A|\chi_B}$ is given in the first column.
    The subsequent columns describe the properties of the configuration:
    the \ac{EOS}, cf.~\cite{Read:2008iy},
    the NS' individual masses $M_{A,B}$,
    the stars' dimensionless spins $\chi_{A,B}$,
    the stars' compactnesses $C_{A,B}$,
    the tidal deformabilities of the stars $\Lambda_{A,B}$,
    the tidal deformability of the binary $\tilde{\Lambda}$,
    the effective dimensionless coupling constant
    $\kappa^T_{\rm eff}$,
    and the merger frequency $f_{\rm mrg}$ in kHz.
    The last three columns give information about the \ac{NR} dataset:
    the initial frequency of the simulation $\mo_0$,
    the residual eccentricity~\cite{Dietrich:2015pxa},
    and the grid resolutions $h$ covering the \ac{NS}.}
\begin{small}
\begin{tabular}{l||cccccccccccc|ccc}
\hline
\hline
  Name & EOS & $M_{A}$ & $M_B$& $\chi_{A}$ & $\chi_{B}$ & $C_A$& $C_B$ & $\Lambda_A$ & $\Lambda_B$ & $\tilde{\Lambda}$ & $\kappa^T_{\rm eff}$ & $f_{\rm mrg} [kHz]$ &  $\mo_0$ & $e [10^{-3}]$  & $\Delta x$  \\
     \hline
\multicolumn{12} {l} {\bfseries equal mass, non-spinning}     \\
\hline
\hline
2B$_{1.35|1.35}^{0.00|0.00}$   & 2B   & 1.3500 & 1.3500 & 0.000 & 0.000 & 0.205 & 0.205 & 127.5 & 127.5 & 127.5 & 23.9  & 2.567 & 0.038 & 7.1  & 0.093 \\
SLy$_{1.38|1.38}^{0.00|0.00}$   & SLy & 1.3750 & 1.3750 & 0.000 & 0.000 & 0.178 & 0.178 & 347.3 & 347.3 & 347.3 & 65.1  & 1.978 & 0.036 & 14.6 & 0.116 \\
SLy$_{1.35|1.35}^{0.00|0.00}$  & SLy  & 1.3500 & 1.3500 & 0.000 & 0.000 & 0.174 & 0.174 & 392.1 & 392.1 & 392.1 & 73.5  & 2.010 & 0.038 & 0.4  & 0.059 \\
H4$_{1.37|1.37}^{0.00|0.00}$   & H4   & 1.3717 & 1.3717 & 0.000 & 0.000 & 0.149 & 0.149 & 1013.4& 1013.4& 1013.4& 190.0 & 1.535 & 0.037 & 0.9  & 0.083 \\
MS1b$_{1.38|1.38}^{0.00|0.00}$  & MS1b& 1.3750 & 1.3750 & 0.000 & 0.000 & 0.144 & 0.144 & 1389.4& 1389.4& 1389.4& 260.5 & 1.416 & 0.035 & 4.0  & 0.116 \\
MS1b$_{1.35|1.35}^{0.00|0.00}$ & MS1b & 1.3500 & 1.3500 & 0.000 & 0.000 & 0.142 & 0.142 & 1536.7& 1536.7& 1536.7& 288.1 & 1.405 & 0.036 & 1.7  & 0.097 \\
\hline
\multicolumn{12} {l} {\bfseries equal mass, spinning}     \\
\hline
\hline
SLy$_{1.35|1.35}^{0.05|0.05}$    & SLy  & 1.3502 & 1.3502 & +0.052 & +0.052 & 0.174 & 0.174 & 392.0 & 392.0 & 392.0& 73.5  & 2.025 & 0.038 & 0.4 & 0.078 \\
SLy$_{1.35|1.35}^{0.11|0.11}$    & SLy  & 1.3506 & 1.3506 & +0.106 & +0.106 & 0.174 & 0.174 & 391.0 & 391.0 & 391.0& 73.5  & 2.048 & 0.038 & 0.7 & 0.078 \\
H4$_{1.37|1.37}^{0.14|0.14}$     & H4   & 1.3726 & 1.3726 & +0.141 & +0.141 & 0.149 & 0.149 & 1009.1& 1009.1&1009.1& 189.2 & 1.605 & 0.037 & 0.4 & 0.083 \\
MS1b$_{1.35|1.35}^{-0.10|-0.10}$ & MS1b & 1.3504 & 1.3504 & -0.099 & -0.099 & 0.142 & 0.142 & 1534.5& 1534.5&1534.5& 287.7 & 1.323 & 0.036 & 1.8 & 0.097 \\
MS1b$_{1.35|1.35}^{0.10|0.10}$   & MS1b & 1.3504 & 1.3504 & +0.099 & +0.099 & 0.142 & 0.142 & 1534.5& 1534.5&1534.5& 287.7 & 1.442 & 0.036 & 1.9 & 0.097 \\
MS1b$_{1.35|1.35}^{0.15|0.15}$   & MS1b & 1.3509 & 1.3509 & +0.149 & +0.149 & 0.142 & 0.142 & 1531.8& 1531.8&1531.8& 287.2 & 1.456 & 0.036 & 1.8 & 0.097 \\
\hline
\multicolumn{12} {l} {\bfseries unequal mass, non-spinning}     \\
\hline
\hline
SLy$_{1.53|1.22}^{0.00|0.00}$   & SLy & 1.5274 & 1.2222 & 0.000 & 0.000 & 0.198 & 0.157 & 167.5 & 732.2  & 365.6  & 68.6  & 1.770 & 0.036 & 8.3  & 0.125  \\
SLy$_{1.65|1.10}^{0.00|0.00}$   & SLy & 1.6500 & 1.0979 & 0.000 & 0.000 & 0.215 & 0.142 & 93.6  & 1372.3 & 408.1  & 76.5  & 1.592 & 0.036 & 8.0  & 0.116 \\
SLy$_{1.50|1.00}^{0.00|0.00}$   & SLy & 1.5000 & 1.0000 & 0.000 & 0.000 & 0.194 & 0.129 & 192.3 & 2315.0 & 720.0  & 135.0 & 1.504 & 0.031 & 11.9 & 0.125 \\
MS1b$_{1.53|1.22}^{0.00|0.00}$  & MS1b& 1.5278 & 1.2222 & 0.000 & 0.000 & 0.159 & 0.130 & 779.6 & 2583.2 & 1420.4 & 266.3 & 1.301 & 0.035 & 8.3  & 0.125 \\
MS1b$_{1.65|1.10}^{0.00|0.00}$  & MS1b& 1.6500 & 1.1000 & 0.000 & 0.000 & 0.171 & 0.118 & 505.2 & 4405.9 & 1490.1 & 279.4 & 1.170 & 0.035 & 8.0  & 0.116 \\
MS1b$_{1.500|1.00}^{0.00|0.00}$ & MS1b& 1.5000 & 1.0000 & 0.000 & 0.000 & 0.157 & 0.109 & 866.5 & 7041.6 & 2433.5 & 456.3 & 1.113 & 0.030 & 11.9 & 0.125 \\
     \end{tabular}
     \end{small}
 \label{tab:NRconfigs}
\end{table*}

\subsection{TEOBResumS}

The \TEOBResumS model is an EOB waveform model that is able to generate BBH waveforms
through merger and ringdown, and tidally modified waveforms up to merger.
The model can deal with spin-aligned binaries and is based on several
recent theoretical developments~\cite{Bernuzzi:2012ku,Damour:2014sva,Harms:2014dqa,
Nagar:2015xqa,Nagar:2017jdw,Messina:2018ghh}.

In its more recent version, the model is able to blend together, in resummed
form, tidal and spin effects~\cite{Nagar:TEOBResumS}. Notably, through a suitable
modification of the concept of centrifugal radius introduced in Ref.~\cite{Damour:2014sva},
it is easily possible to incorporate the \ac{EOS} dependent self-spin effects
(or quadrupole-monopole terms~\cite{PhysRevD.57.5287}) within the EOB
Hamiltonian and flux. The current version of \TEOBResumS we are dealing here
does this at LO only, while the EOB extension to NLO order, from the NLO spin-spin
PN results of Ref.~\cite{Levi:2014sba,Bohe:2015ana}, will be done
elsewhere~\cite{DBN:2018} (we recall in this respect that in the BBH sector
NLO spin-spin interaction is already incorporated within \TEOBResumS).

By contrast, we stress that some of the PN waveform approximants that we
discuss below, notably \TaylorFT and \PhenomPNRtidal, do incorporate the
NLO information. We outline the importance of this difference specifically
in Sec.~\ref{sec:checkSS}. Note, however, that the resummation itself makes the
behavior of the self-spin coupling different from the standard PN treatment,
notably making it more attractive during the inspiral~\cite{Nagar:TEOBResumS}.

In addition, the spin-tidal sector of \TEOBResumS differs from the BBH model
recently upgraded in Ref.~\cite{Nagar:2017jdw} in that the effective,
next-to-next-to-next-to-leading order spin-orbit parameter $c_3$ that is
informed by NR simulation is here neglected (i.e. $c_3=0$).\footnote{We anyway 
tested that the
  effect of using the value of $c_3$ informed by BBH simulations~\cite{Nagar:2017jdw}
  is essentially negligible for the range of spins considered here. This is meaningful as
  tidal interaction screens the effects of the high-PN correction yielded by 
$c_3$.}

The \TEOBResumS model has been validated through phase comparisons with
\ac{NR} simulations~\cite{Bernuzzi:2014owa}: 
EOB and NR waveforms are found to agree well in most regions of the \ac{BNS} parameter space; 
slightly larger dephasing are found for models with large values of
the tidal parameter (e.g., based on the MS1b \ac{EOS}), suggesting that some
improvements in the model are still needed. We finally recall that all EOB
waveforms generated here were obtained using post-post-circular initial
data consistently generalized to the spinning-tidal case~\cite{Damour:2012ky,Nagar:TEOBResumS}.

\subsection{Numerical relativity waveforms}
The \ac{NR} simulations used for validation of the
\NRtidal waveform models have been computed with the \bam
code, with details given
in~\cite{Brugmann:2008zz,Thierfelder:2011yi,Dietrich:2015iva,Bernuzzi:2016pie}.
For all simulations, we employ the Z4c scheme~\cite{Bernuzzi:2009ex,Hilditch:2012fp}
for the spacetime evolution and the 1+log and
gamma-driver conditions ~\cite{Bona:1994a,Alcubierre:2002kk,vanMeter:2006vi,
Campanelli:2005dd,Baker:2005vv} for the gauge system.
Finite difference stencils are used for the spatial discretization of
the spacetime and high resolution shock-capturing methods
for the hydrodynamics part are applied.
We summarize the configurations employed in this work in Tab.~\ref{tab:NRconfigs}.
Overall, we use $18$ different physical configurations.
The setups span $4$ different \acp{EOS}; in particular, 2B and MS1b were chosen as relatively extreme cases to test the performance across the
\ac{EOS} parameter space, because both 2B and MS1b are almost ruled out after the multi-messenger observation
of GW170817~\cite{Radice:2017lry}.
We also test mass ratios up to $q =1.5$. While such mass ratios are possible based on
binary evolution models~\cite{Dominik:2012kk,Dietrich:2015pxa}, no observed \ac{BNS} system
has such a large mass ratio~\cite{2012ARNPS..62..485L}.
Similarly, to date no \ac{NS} in a \ac{BNS} system has
a dimensionless spin larger than $\sim 0.05$,
but nevertheless we consider values up to $\chi = 0.15$.
Let us emphasize that while the considered configurations cover most of the
\ac{BNS} parameter space which we expect to detect, 
for parameter estimation from GW observations
waveforms in even larger regions need to be 
evaluated~\footnote{Although potentially irrelevant for real physical systems, 
let us point out that for some parameter combinations, e.g., large deformabilities
$\Lambda \sim 5000$ or almost extremal-antialigned spin
$\chi_A= \chi_B = -0.99$ unphysical features in the
waveforms can be seen. Therefore, we emphasize that interpretation of results of 
parameter estimation pipelines in extreme (unphysical) corners of the 
BNS parameters space need to be taken with care and requires special attention.}.
Consequently, one important goal for \ac{NR} simulations of BNSs is to access
unexplored regions in terms of masses, mass-ratios, and spins.
\subsection{Hybrid construction}
\begin{figure}[t]
\includegraphics[width=0.5\textwidth]{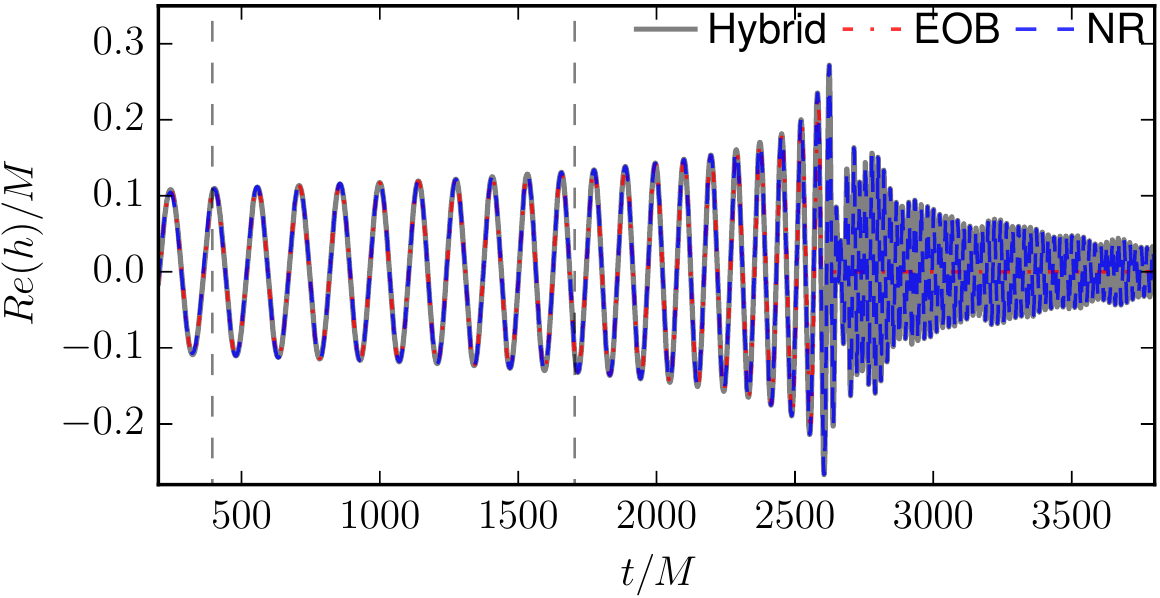}
\caption{Hybridization of the SLy$_{1.35|1.35}^{0.11|0.11}$
configuration with the \TEOBResumS model. The alignment interval
is marked by vertical dashed lines. The \TEOBResumS \ac{EOB} waveform
is shown as a red dot-dashed curve and the \ac{NR} waveform as a blue dashed curve.
The final hybrid combines the long inspiral from the \ac{EOB} waveform,
which includes several hundred cycles (not shown in the figure), and the
late inspiral, as well as the post-merger phase of the \ac{NR} waveform. }
\label{fig:hybrid}
\end{figure}

The procedure for hybridizing the \ac{EOB} and \ac{NR} waveforms is as follows.
We align the \ac{EOB} and \ac{NR} waveforms, which employ the same binary parameters, by minimizing
\begin{equation}
\mathcal{I}(\delta t, \delta \phi) = \int_{t_i}^{t_f}dt|\phi_{NR}(t) -
\phi_{\rm EOB}(t+\delta t) + \delta \phi|^2 \label{eq:alignment}
\end{equation}
over the frequency interval $I_{\mo}=[\mo_i,\mo_f]=[0.04,0.06]$.
Once the waveforms are aligned, we perform a smooth transition from the \ac{EOB} data
to the \ac{NR} data within $I_{\mo}$:
\begin{equation}
h_{\rm hyb}(t) = \left\{
 \begin{array}{lr}
   h_{\rm EOB} & : \mo \le \mo_i\\
   h_{\rm NR}H(t) + h_{\rm EOB} [1- H(t)] & : \mo_i \le \mo \le \mo_f\\
   h_{\rm NR} & : \mo \ge \mo_f
 \end{array}
\right.
\end{equation}
with the Hann window function
\begin{equation}
H(t) := \frac{1}{2} \bigg[ 1 - \cos \bigg( \pi \frac{t-t_i}{t_f - t_i} \bigg) \bigg],
\end{equation}
with $t_i,t_f$ denoting the times corresponding to $\mo_i,\mo_f$,
cf.~\cite{Hotokezaka:2016bzh}.
In Fig.~\ref{fig:hybrid} we present, as an example, the hybrid
construction for the SLy$_{1.35|1.35}^{0.11|0.11}$ configuration, with
the alignment interval marked by vertical dashed lines.

\section{Validation of frequency domain model}
\label{sec:FDchecks}

\subsection{Mismatch computation} \label{sec:mismatches}

To quantify the performance of the \NRtidal approximants, we compute
the mismatch
\begin{equation}
\bar{F} = 1 - \max_{\phi_c,t_c} \frac{(h_1(\phi_c,t_c)|h_2)}{\sqrt{(h_1|h_1)(h_2|h_2)}}\,,
\label{eq:mismatch}
\end{equation}
where $\phi_c,t_c$ are an arbitrary phase and time shift, between the
approximants themselves and the hybrid waveforms constructed in
Sec.~\ref{sec:nrhybrids}.  The noise-weighted overlap is defined as
\begin{equation}
 (h_1 | h_2) = 4 \Re \int_{f_{\rm min}}^{f_{\rm
 max}} \frac{\tilde{h}_1(f) \tilde{h}_2(f)}{S_n(f)} \text{d} f \ .
\end{equation}
$S_n(f)$ gives the spectral density of the detector
noise. We used the Advanced LIGO zero-detuning, high-power (\verb#ZERO_DET_high_P#) noise curve
of~\cite{Sn:advLIGO} for our analysis.
In general, the value of $\bar{F}$ indicates the loss in signal-to-noise ratio
(squared) when the waveforms are aligned in time and phase.
Template banks are usually constructed such that the maximum value of
$\bar{F}$ across the bank is $0.03$.
Although it is impossible to relate a mismatch directly to the bias obtained in
parameter estimation, it is in general a good measure of the performance
of a particular waveform approximant.

\subsubsection{Variable $f_{\rm max}$}

\begin{figure*}[t]
\includegraphics[width=0.98\textwidth]{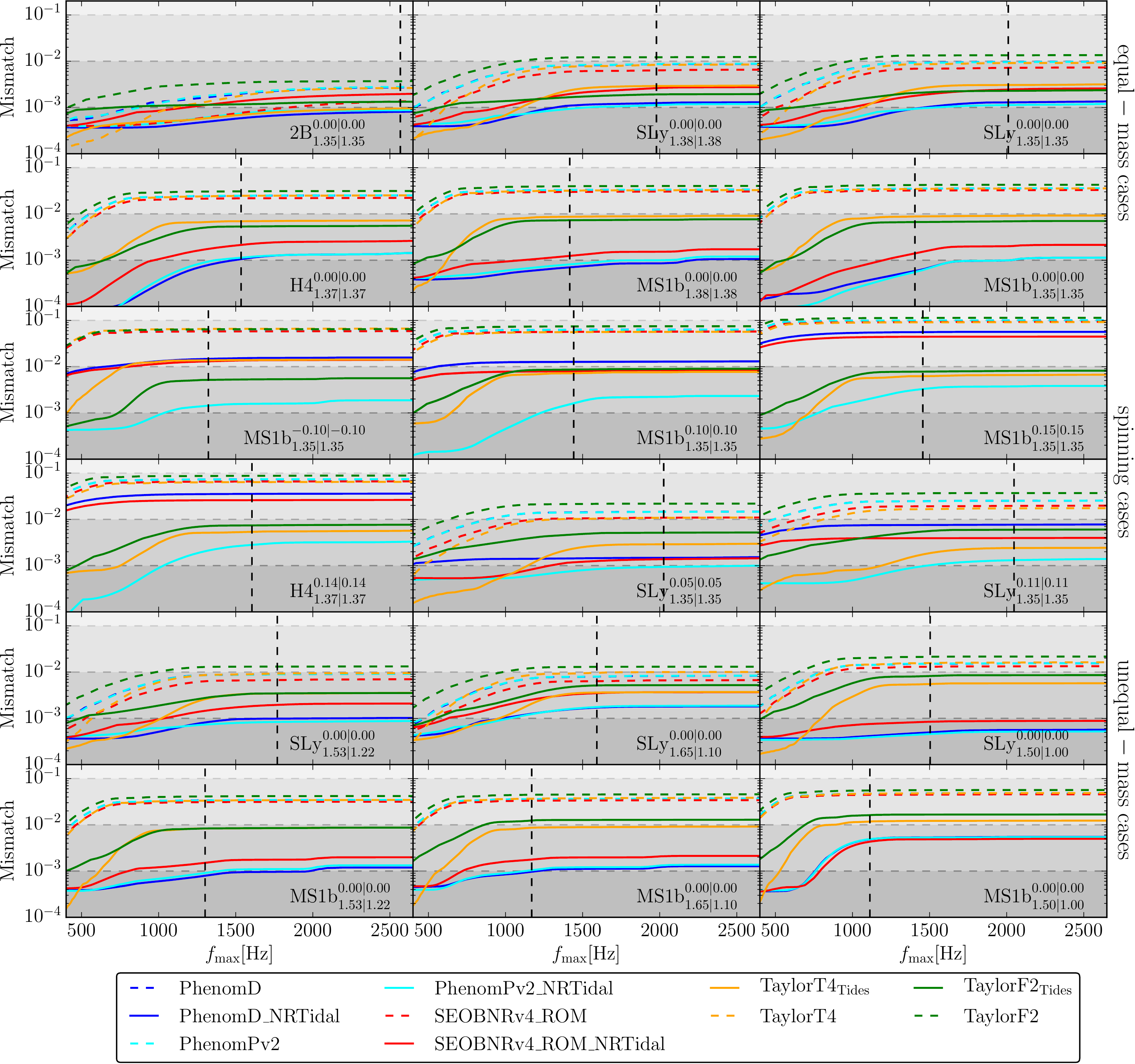}
\caption{
Mismatches between the tidal approximants presented in this paper and the
hybrid waveforms based on the configurations listed in Table~\ref{tab:NRconfigs}.
Mismatches are computed following Eq.~\eqref{eq:mismatch},
where we set $f_{\rm min}=30{\rm Hz}$ and vary the maximum frequency $f_{\rm max}$.
A black vertical dashed line marks the frequency corresponding to the
moment of merger $f_{\rm mrg}$. Note that for the analysis of
GW170817 in~\cite{TheLIGOScientific:2017qsa}, a maximum frequency of $2048{\rm Hz}$ was  employed.
The naming convention of the individual panels refers to the setup described in Tab.~\ref{tab:NRconfigs}, namely: ${\rm EOS}_{M_A|M_B}^{\chi_A|\chi_B}$.
Regarding the waveform approximants, results for all models that do not include tidal effects are marked as dashed lines,
while solid lines refer to waveform models including tidal effects. The color coding is as follows:
\TaylorFT (green), \TaylorTT (orange), \SEOBNRROMNRtidal (red), \PhenomDNRtidal (blue), \PhenomPNRtidal (cyan).
Overall we find that the \PhenomPNRtidal model performs best. In particular, this model is advantageous
for spinning configurations.}
\label{fig:mismatch:fmax}
\end{figure*}

In Fig.~\ref{fig:mismatch:fmax} we report the mismatch between the
proposed model approximants and the hybrid waveforms constructed
in Sec.~\ref{sec:nrhybrids}. In addition to the \NRtidal models
and their underlying point-mass baselines, we also explore the
performance of \ac{PN} based models, in particular the
\TaylorT and \TaylorF approximants
(see e.g.~\cite{Damour:1997ub,Damour:2000gg,Damour:2000zb,PhysRevD.80.084043}).
To set the stage, we recall that we use 3.5PN-accurate
expressions for the non-spinning part of the phase as
well as for the spin-orbit terms~\cite{0264-9381-30-13-135009}.
Up to 3PN-accurate, \ac{EOS}-dependent, self-spin
terms~\cite{PhysRevD.71.124043, PhysRevD.79.104023,0264-9381-32-19-195010},
that are essential for a conceptually meaningful comparison
with the \TEOBResumS-hybrid waveforms, are included in both
\TaylorFT and \TaylorTT approximants. For what concerns
the tidal sector, while \TaylorTT only incorporates the LO and NLO
tidal corrections (i.e. corresponding to a 5PN and 6PN terms),
in \TaylorF we also included the 6.5, 7 and 7.5PN tidal terms
as deduced by Taylor-expanding the tidal EOB model
in Ref.~\cite{Damour:2012yf}.

Results for all approximants incorporating tidal effects are shown with solid lines:
\TaylorFT (green), \TaylorTT (orange), \SEOBNRROMNRtidal (red), \PhenomDNRtidal (blue), \PhenomPNRtidal (cyan).
Results for the corresponding approximants without tidal effects are shown with dashed lines.
The mismatches in Fig.~\ref{fig:mismatch:fmax} are computed from $f_{\rm min}=30$Hz up
to a variable maximum frequency $f_{\rm max}$. We mark the merger frequency extracted
from the \ac{NR} simulations with a vertical, black dashed line.
Let us discuss the different datasets separately.

\paragraph*{Non-spinning, equal-mass configurations:}
While for small tidal deformability waveform models not including tidal
effects also achieve mismatches smaller $5\times 10^{-3}$, e.g.~${\rm 2B}_{1.35|1.35}^{0.00|0.00}$,
this is not true for increasing tidal deformability (left to right).
For stiff \acp{EOS} waveform models not including tidal effects are
inaccurate and mismatches can increase more than an order of magnitude
compared to \NRtidal models,
e.g.~${\rm MS1b}_{1.35|1.35}^{0.00|0.00}$.
Furthermore, mismatches between \TaylorFT and the hybrid waveforms increase with
increasing tidal effects (large values of $\Lambda$).
Validating the performance of the \NRtidal models among each other,
we find that \PhenomDNRtidal and \PhenomPNRtidal tend to
approximate the hybrids slightly better than the \SEOBNRROMNRtidal approximant,
but differences are small.

\paragraph*{Spinning configurations:}
For spinning configurations, one finds that
the value of the mismatches delivered by non-tidal models is generally
unacceptably large ($\gtrsim 1\%$), and it is found to increase with the
spin value (see e.g. the ${\rm MS1b}_{1.35|1.35}$ or ${\rm SLy}_{1.35|1.35}$
configurations in the third and fourth row of Fig.~\eqref{fig:mismatch:fmax}).
The inclusion of \ac{EOS}-dependent effects (both tidal and self-spin ones) is
able to lower the mismatches to an acceptable level. 
Furthermore, we find that, since the various matter-dependent effects are included
in \TaylorFT and \TaylorTT, one also obtains an acceptable agreement 
($<1\%$ during the inspiral) with the hybrid waveform. 
As expected, the smallest values ($\simeq 0.1\%$) are obtained when 
the spins are small, and the \ac{EOS} is soft, e.g.~${\rm SLy}_{1.35|1.35}^{0.05|0.05}$. 
This is not surprising since, for example for \TaylorTT, 
it is known that the the point-mass (nontidal), nonspinning baseline is 
just by chance, especially reliable in the equal-mass, nonspinning case 
(see e.g. Ref.~\cite{Boyle:2007ft,Damour:2007yf}),
although it has the property of generically underestimating the tidal
forces~\cite{Baiotti:2010xh};
Furthermore, both \PhenomDNRtidal and \SEOBNRROMNRtidal exceed the
$1\%$ limit during the inspiral except for the ${\rm SLy}_{1.35|1.35}$
configurations where only the small-spin configuration
${\rm SLy}_{1.35|1.35}^{0.05|0.05}$ is around the $0.1\%$ level.
Interestingly, once the \PhenomDNRtidal model is completed by
the \ac{EOS}-dependent self-spin terms, as it is done in the \PhenomPNRtidal,
the mismatches drop at, or below, the $10^{-3}$ level (i.e., by up to more than an order of magnitude)
for all configurations considered in the two central rows of Fig.~\ref{fig:mismatch:fmax}.
This suggests that the \PhenomPNRtidal is very effective in representing 
the LO self-spin terms incorporated within the \TEOBResumS model.
Overall, our analysis shows that for sufficiently stiff \acp{EOS}, even relatively small spin
magnitudes $(\sim 0.1)$ are sufficient to have an effect on mismatches between
long signals starting at $f_{\rm min}=30$Hz. By contrast, 
note that the mismatches are less affected by the self-spin effects for
${\rm SLy}_{1.35|1.35}^{0.05|0.05}$. 

On the basis of this analysis, we can state that \PhenomPNRtidal (or similarly \PhenomDNRtidal
once augmented with the \ac{EOS}-dependent self-spin effects) delivers the closes matches to
the EOB-NR hybrid waveforms and, thus, it is preferred with respect to the other approximants
currently implemented in LAL.

\paragraph*{Unequal mass configurations:}
For unequal mass, nonspinning, binaries the importance of the inclusion of tidal effects
is also evident. Furthermore, we see that the \TaylorFT model has the largest mismatch
among all tidal approximants. Likewise the equal-mass, nonspinning case mentioned above,
\PhenomPNRtidal and \PhenomDNRtidal are essentially equivalent.

\subsubsection{Variable $f_{\rm min}$}

\begin{figure*}[t]
\includegraphics[width=0.98\textwidth]{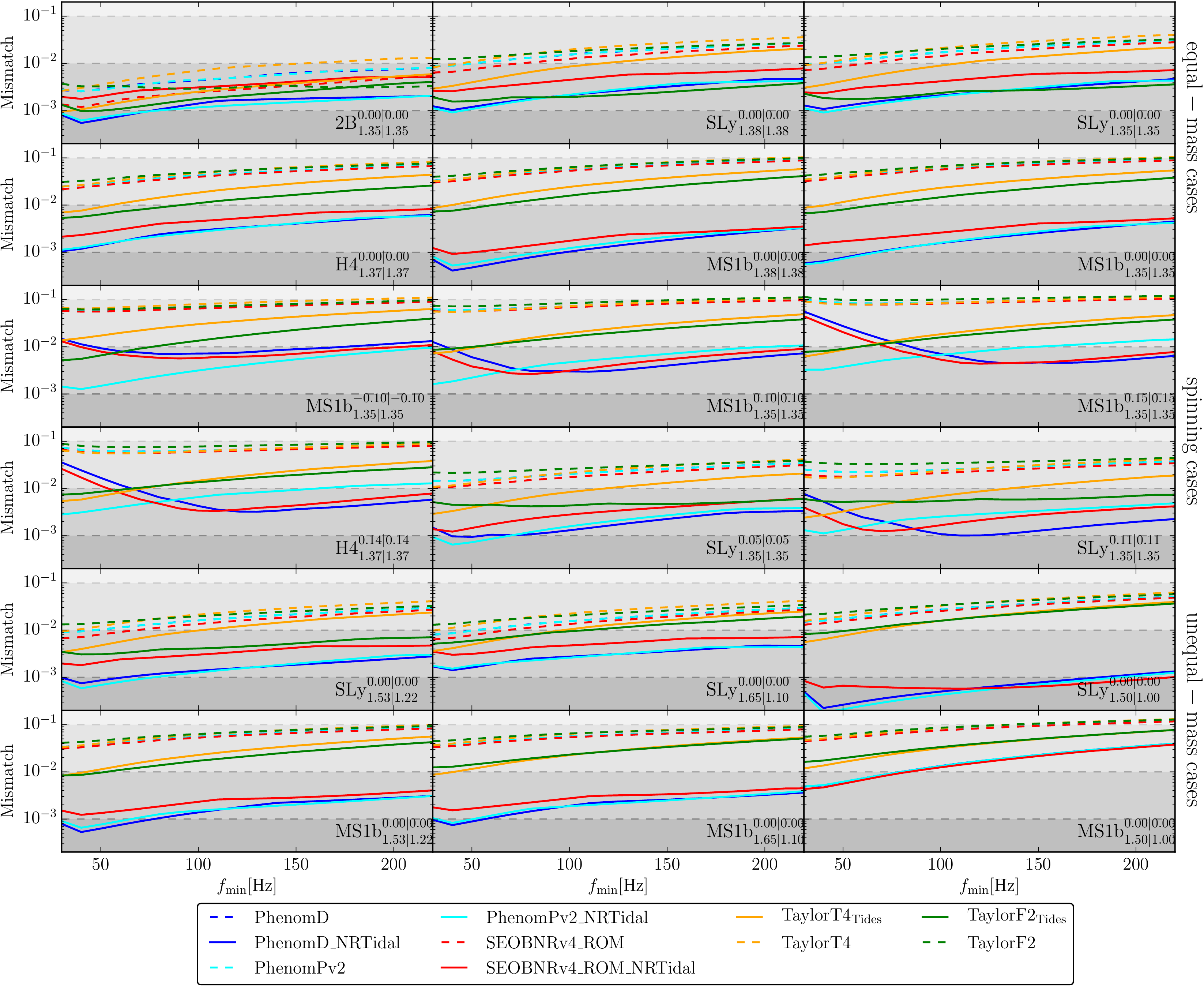}
\caption{
Same configurations as Fig.~\ref{fig:mismatch:fmax}, but with mismatches computed
varying the minimum frequency $f_{\rm min}$ in Eq.~\eqref{eq:mismatch} The maximum frequency $f_{\rm max}$
is kept fixed and equal to the merger frequency
of the \ac{BNS} configurations (i.e., the values of the vertical dashed lines
in the panels of Fig.~\ref{fig:mismatch:fmax}).
}
\label{fig:mismatch:fmin}
\end{figure*}

In addition to our previous investigation, we now study the effect of varying the minimum
frequency $f_{\rm min}$ while setting the maximum frequency $f_{\rm max}$ equal to the merger frequency
of the \ac{BNS} configuration $f_{\rm mrg}$ (dashed black lines in Fig.~\ref{fig:mismatch:fmax}).
As $f_{\rm min}$ increases, we will be looking at a signal with decreasing length, putting
more emphasis on the late-inspiral, which is generally harder to model
since gravitational forces and tidal effects are stronger.
Therefore, we expect that as $f_{\rm min}$ increases, so will the mismatches.
Figure~\ref{fig:mismatch:fmin} summarizes our results, detailed below.

\paragraph*{Non-spinning, equal-mass configurations:}
In the absence of spins, we can distinguish three different regimes.
(i) When tidal effects are small, cf.~${\rm 2B}_{1.35|1.35}^{0.00|0.00}$,
even non-tidal approximants have mismatches below $10^{-2}$.
Nevertheless, tidal approximants have comparatively smaller mismatches.
(ii) With increasing values of $\Lambda$, non-tidal models fail
to approximate the hybrid waveforms, with mismatches that easily exceed
the $10^{-2}$ threshold.
(iii) For even larger values of $\Lambda$, the mismatches with the
\TaylorFT  approximant are about an order of magnitude worse than
the various \NRtidal models, notably even exceeding the $10^{-2}$ level.

\paragraph*{Spinning configurations:}
For spinning configurations, all non-tidal models have
mismatches of the order of $10^{-2}$--$10^{-1}$ for frequencies
$f_{\rm min}=200{\rm Hz}$.
We find that generally the \PhenomPNRtidal model performs best
for frequencies below $f_{\rm min}<100{\rm Hz}$. For some setups,
e.g.~${\rm MS1b}_{1.35|1.35}^{0.10|0.10}$, ${\rm SLy}_{1.35|1.35}^{0.11|0.11}$,
the mismatches with respect to \SEOBNRROMNRtidal and \PhenomDNRtidal decrease
as $f_{\rm min}$ increases and have a minimum
around $100$--$150{\rm Hz}$. We suggest that this effect is again caused
by the \ac{EOS} dependent spin-induced quadrupole term which is neglected in 
the LALSuite implementation of \SEOBNRROMNRtidal and \PhenomDNRtidal.
Once $f_{\rm min}$ is increased, the signal used for the mismatch computation becomes shorter,
which in turn suppresses the error introduced by neglecting the quadrupole-monopole term.
However, later in the evolution the mismatches increase again as for all other models
due to inaccuracies in the description of the strong-gravity regime.

\paragraph*{Unequal mass configurations:}
For unequal masses, \TaylorFT produces mismatches
about an order of magnitude worse than the \NRtidal models.
However, we also find that for high mass ratios and large tidal effects the
\NRtidal approximants become less accurate.
In this case, e.g.~${\rm MS1b}_{1.50|1.00}^{0.00|0.00}$, mismatches remain below $10^{-2}$
only if $f_{\rm min}$ is smaller than $100{\rm Hz}$.

\subsection{Dephasing}

In addition to mismatches, we also use phase differences computed in the
frequency-domain between the hybrid waveforms and the various waveform
approximants as a way to judge the performance of the waveform models.
Phase differences may provide
information that is complementary to the mismatch study presented in
Sec.~\ref{sec:mismatches} for the following reasons. First, mismatches weight
waveform differences according to the assumed noise spectral density and the
signal amplitudes. For \ac{BNS} signals in Advanced LIGO, this means
that mismatches are less sensitive to waveform disagreements in the late
inspiral. However, we are interested in assessing the
quality of the models in this regime as well.

Second, the matches we calculate
are optimized over a relative time and phase shift between hybrids and models
that are subject to the same amplitude and noise weighting as described above.
We now apply an independent time and phase alignment in the frequency domain
by minimizing the average square difference between the hybrid's phase
[$\Psi_1(f)$] and the model's phase [$\Psi_2(f)$],
\begin{align} \label{eq:square_phase_diff}
\Delta \Psi_{\mathbb L_2}^2 = \min_{t_0, \Psi_0}
\int_{f_{\rm min}}^{f_{\rm max}} \frac{(\Psi_{1} - \Psi_{2} + 2\pi f t_0
+ \Psi_0)^2}{f_{\rm max} - f_{\rm min}} df.
\end{align}
For the optimal values of $t_0$ and $\Psi_0$ in Eq.~\eqref{eq:square_phase_diff},
we additionally analyze the maximal value of the phase difference, $\Delta
\Psi_{\max}$ [i.e., the maximum of the square root of the numerator in
Eq.~\eqref{eq:square_phase_diff}].
Finally, we can localize the origin of the
observed dephasing in an \emph{alignment-independent way} by analyzing the
second phase derivative (see the discussion below).

For a broadband alignment from $f_{\rm min} = 50\,{\rm Hz}$ to $f_{\rm max} =
f_{\rm mrg}$, we find that the models augmented with \ac{NR}-tuned tidal phase
corrections exhibit phase differences $\Delta
\Psi_{\mathbb L_2} \leq 0.5$ for all hybrids except ${\rm
MS1b}_{1.50|1.00}^{0.00|0.00}$ for which $\Delta
\Psi_{\mathbb L_2} \approx 0.7$. The maximal dephasing is $\Delta
\Psi_{\rm max} < 1.3$ for all hybrids. These values are
consistently smaller than the results for the respective point-particle
baseline models. In particular, without \ac{NR}-tuned tidal corrections, both
$\Delta
\Psi_{\mathbb L_2}$ and $\Delta \Psi_{\rm max}$ increase by a factor of 2--9
for soft \acp{EOS} and factors of 7--23  for stiff \acp{EOS}. We note
that tidal \ac{PN} approximants show a smaller improvement compared to their
respective point-particle description, and in the case of the 2B \ac{EOS}, the
tidal \ac{PN} models even have a larger dephasing than their point-particle
counterparts. In general, pure \ac{PN} approximants perform worse or similar at
best compared to the \ac{NR}-tuned tidal models.

Interestingly, the broadband results discussed above do not
hold universally when we align only at lower frequencies,
e.g., $(f_{\rm min}, f_{\max}) = (50, 500)\,{\rm Hz}$. In this interval,
\NRtidal
approximants remain superior to their point-particle counterparts. However, for spinning configurations, \ac{PN} tidal approximants
now perform typically better than or as well as \PhenomDNRtidal and \SEOBNRROMNRtidal.
\PhenomPNRtidal consistently shows the smallest dephasing from all hybrids. As
discussed for the mismatch results, we attribute this mainly to spin-dependent
quadrupole terms that are included in \PhenomPNRtidal and both \ac{PN}
approximants, but are missing in the other two \NRtidal approximants in the current LALSuite implementation.

As a final corroboration of our results, we now
localize the origin of the dephasing between different waveform models. Doing
this based on the frequency-domain phase $\Psi(f)$ is ambiguous due to the
freedom of time and phase shifting each waveform. However, the second derivate
of $\Psi$ removes all degrees of freedom associated with time and phase shifts
(as they manifest themselves as a linear function in the frequency domain). For
signals with slowly varying amplitude, the stationary-phase-approximation
allows us to identify $d \Psi / df$ as proportional to the time at which each
frequency is realized. Consequently,
\begin{equation}
 \tau (f) = \frac{d^2 \Psi(f)}{df^2}  \label{eq:tau}
\end{equation}
may be interpreted as the time the signal spends per unit frequency in the inspiral.
(The units of
$\tau$ are ${\rm s}^2$ or equivalently ${\rm s} / {\rm Hz}$.)

\begin{figure}
 \includegraphics[width=\columnwidth]{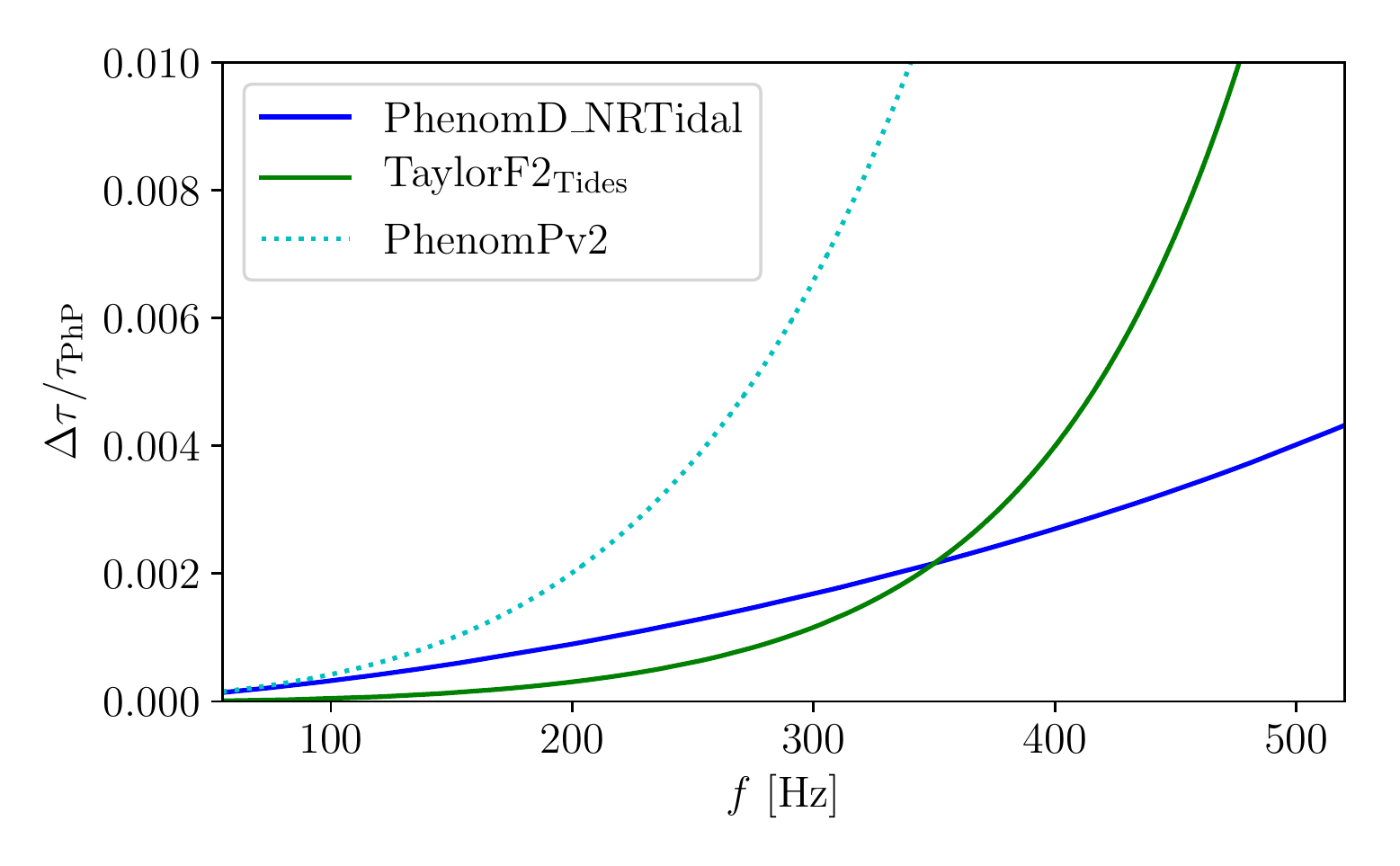}
 \caption{Time spent by GW signals per unit frequency [see Eq.~\eqref{eq:tau} and
surrounding discussion], for different waveform models and
shown here as relative differences to \PhenomPNRtidal for the ${\rm
MS1b}_{1.35|1.35}^{0.15|0.15}$ case. The crossing of \TaylorFT and
\PhenomDNRtidal separates the frequency regimes in which either the
spin-dependent quadrupole term (low frequencies) or the improved tidal
corrections (high frequencies) are dominant.}
\label{fig:phase_derivatives}
\end{figure}

Figure~\ref{fig:phase_derivatives} shows the relative differences in $\tau$
between frequency-domain approximants and our best-performing model,
\PhenomPNRtidal for ${\rm MS1b}_{1.35|1.35}^{0.15|0.15}$.
Not surprisingly, neglecting tidal effects completely leads to a visible
disagreement across all frequencies we show (cf.\ \PhenomP curve).
Interestingly, below $\sim350\, {\rm Hz}$, we see that \TaylorFT agrees
better with \PhenomPNRtidal than \PhenomDNRtidal does. We stress again that this is
due to missing, spin-dependent quadrupole terms in \PhenomDNRtidal. Above
$\sim350\, {\rm Hz}$, however, the improved NR-tuned tidal corrections are
more important than the quadrupole terms, which leads to a rapid decline in
accuracy in \TaylorFT, while \PhenomDNRtidal agrees better with \PhenomPNRtidal
at those frequencies.

\section{Comparison with time domain waveforms}
\label{sec:TDchecks}

In the following we also want to test the performance of the waveform models in the time domain.
For this purpose we compute via inverse Fourier transformation
the waveform strain $h(t)$.
As representative cases, we show the equal-mass non-spinning ${\rm SLy}_{1.35|1.35}^{0.00|0.00}$,
the equal-mass spinning ${\rm H4}_{1.37|1.37}^{0.14|0.14}$, and the non-equal mass non-spinning
${\rm MS1b}_{1.65|1.10}^{0.00|0.00}$,
but similar results are obtained for other configurations.
We focus on two different comparisons:
(i) We align waveforms computed from different approximants with the hybrid waveform
several hundred orbits before the actual merger.
At this stage one could expect that all models allow a reasonable prescription
of the binary dynamics and the alignment procedure is justified.
(ii) We align the waveforms obtained from different waveform approximants with the hybrid waveforms 
about $15$ orbits before the merger. At this time tidal effects influence the binary dynamics
and the alignment procedure using non-tidal waveforms
is purely artificial (see discussion below).

Both time-domain alignment procedures are different from the ones carried out in the frequency domain.
While in the frequency domain phase difference
and mismatch computations are usually computed over the entire frequency interval,
in the time domain we aim at studying the accumulation of errors during the binary evolution.

As a final check we also compare the precessing \PhenomPNRtidal model with a
precessing \ac{NR} waveform of~\cite{Dietrich:2017xqb}.
While this comparison is limited to the last 15 orbits before merger, it provides a first qualitative
assessment of the accuracy of the \PhenomPNRtidal model for precessing systems.

\subsection{Waveform alignment in the early inspiral}

\begin{figure*}[t]
\includegraphics[width=0.98\textwidth]{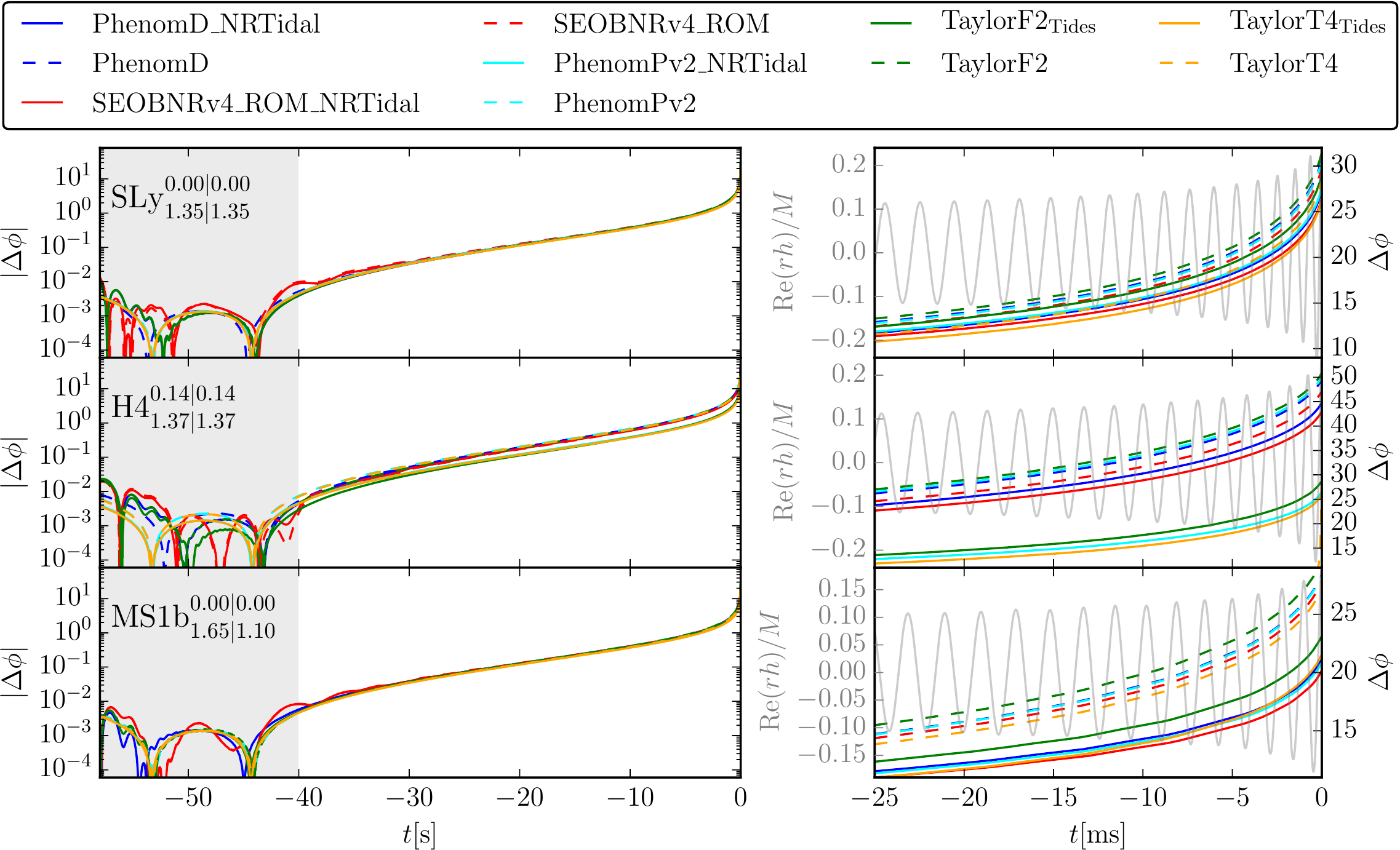}
\caption{
Time domain dephasing $\Delta \phi$ between waveform models and hybrids for
${\rm SLy}_{1.35|1.35}^{0.00|0.00}$ (top panels),
${\rm H4}_{1.37|1.37}^{0.14|0.14}$ (middle panels),
${\rm MS1b}_{1.65|1.10}^{0.00|0.00}$ (bottom panels).
Dashed lines refer to dephasings obtained from waveform models not incorporating
tidal effects, while solid lines include tidal effects (the color coding is the same as in Fig.~\ref{fig:mismatch:fmax}).
The hybrid waveforms and model waveforms are aligned according
to Eq.~\eqref{eq:alignment} in the time interval $t \in [-58~{\rm s},-40~{\rm s}]$
before the merger, cf.~gray shaded region.
The right panels show only the last $25~{\rm ms}$ before merger, i.e., the last few gravitational wave cycles, and
the real part of the hybrid waveforms is shown in gray for a better
visual interpretation.}
\label{fig:time_domain}
\end{figure*}

We consider the last $58~{\rm s}$ before the merger. During this time the
\acp{NS} complete $\sim 1400$ orbits, where the exact number depends on the
configuration details.
The time-domain dephasing $\Delta \phi =  \phi_{\rm hybrid} - \phi_{\rm model}$
is shown on a logarithmic scale in the left panels of Fig.~\ref{fig:time_domain} and
on a linear scale focusing on the last few orbits in the right panels.
We align the waveforms in the time interval $t \in [-58~{\rm s},-40~{\rm s}]$, where
$t=0~{\rm s}$ marks the end of the inspiral of the hybrid.
The color coding of the waveform approximants is
identical to the previous figures.
We will now discuss each individual waveform separately.

\paragraph*{Non-spinning, equal mass} (${\rm SLy}_{1.35|1.35}^{0.00|0.00}$):
Over the time interval considered, the \acp{NS} perform $1388$ orbits,
i.e., the full signal contains $2775$~\ac{GW} cycles (a total of
$17434~{\rm rad}$).

The phase differences between all models and the hybrid waveform
is below $30~{\rm rad}$; as shown in the right, top panel of Fig.~\ref{fig:time_domain},
most of the phase difference is accumulated during the
last $\sim 15$ \ac{GW} cycles. For all models considered the difference between tidal
and non-tidal waveforms is small: $\lesssim$ 1.5 full \ac{GW} cycles.
While almost all waveform models incorporating tidal effects perform
equally well, the \TaylorFT model has the worst performance, while
\SEOBNRROMNRtidal and \TaylorTT perform best.

\paragraph*{Spinning, equal mass} (${\rm H4}_{1.37|1.37}^{0.14|0.14}$):
The overall phase accumulated in the total time interval is about $17272~{\rm rad}$,
which corresponds to $1378$ orbits, or a total of $2750$ \ac{GW} cycles before the
merger of the two stars.
Due to the larger effective tidal coupling constant ---
$\kappa_{\rm eff}^T=189.2$, as opposed to $\kappa_{\rm eff}^T=73.5$ for the ${\rm SLy}_{1.35|1.35}^{0.00|0.00}$ case ---
we also find larger phase differences between \PhenomD, \SEOBNRROM and their
\NRtidal counterparts. However, the main phase difference caused by matter effects
comes from the spin induced quadrupole moment which effects the dynamics
significantly earlier than the tidal contributions modeled in \PhenomDNRtidal and
\SEOBNRROMNRtidal. Therefore, phase differences
of about $20~\rm{rad}$ between \PhenomP and \PhenomPNRtidal are obtained.
Indeed the effect of the spin induced quadrupole contribution is already visible about $30$~s
before the merger, cf.~left, middle panel.
Consequently, for an accurate description of the entire \ac{GW} signal for spinning
NSs we do emphasize again the importance of incorporating the EOS dependent
contributions which are coupled to the star's intrinsic rotation.

\paragraph*{Non-spinning, unequal mass} (${\rm MS1b}_{1.65|1.10}^{0.00|0.00}$):
The ${\rm MS1b}_{1.65|1.10}^{0.00|0.00}$ configuration accumulates
$17642~{\rm rad}$ before the moment of merger, which corresponds to
1390 orbits, i.e.~2780 \ac{GW} cycles.
Overall, phase differences between models that incorporate tidal effects
and those that do not are of the order of $10~{\rm rad}$.
Comparing the performance of tidal waveform models, the \TaylorFT model's performance is
again the worst. The best performance is obtained by the \SEOBNRROMNRtidal model,
with a phase difference of about $20~{\rm rad}$ compared to the hybrid waveform.
Since this configuration contains irrotational \acp{NS}, no phase difference is
visible in the early inspiral between the different approximants,
see bottom left panel.

\subsection{Waveform alignment in the strong-field regime}

\begin{figure*}[t]
\includegraphics[width=0.98\textwidth]{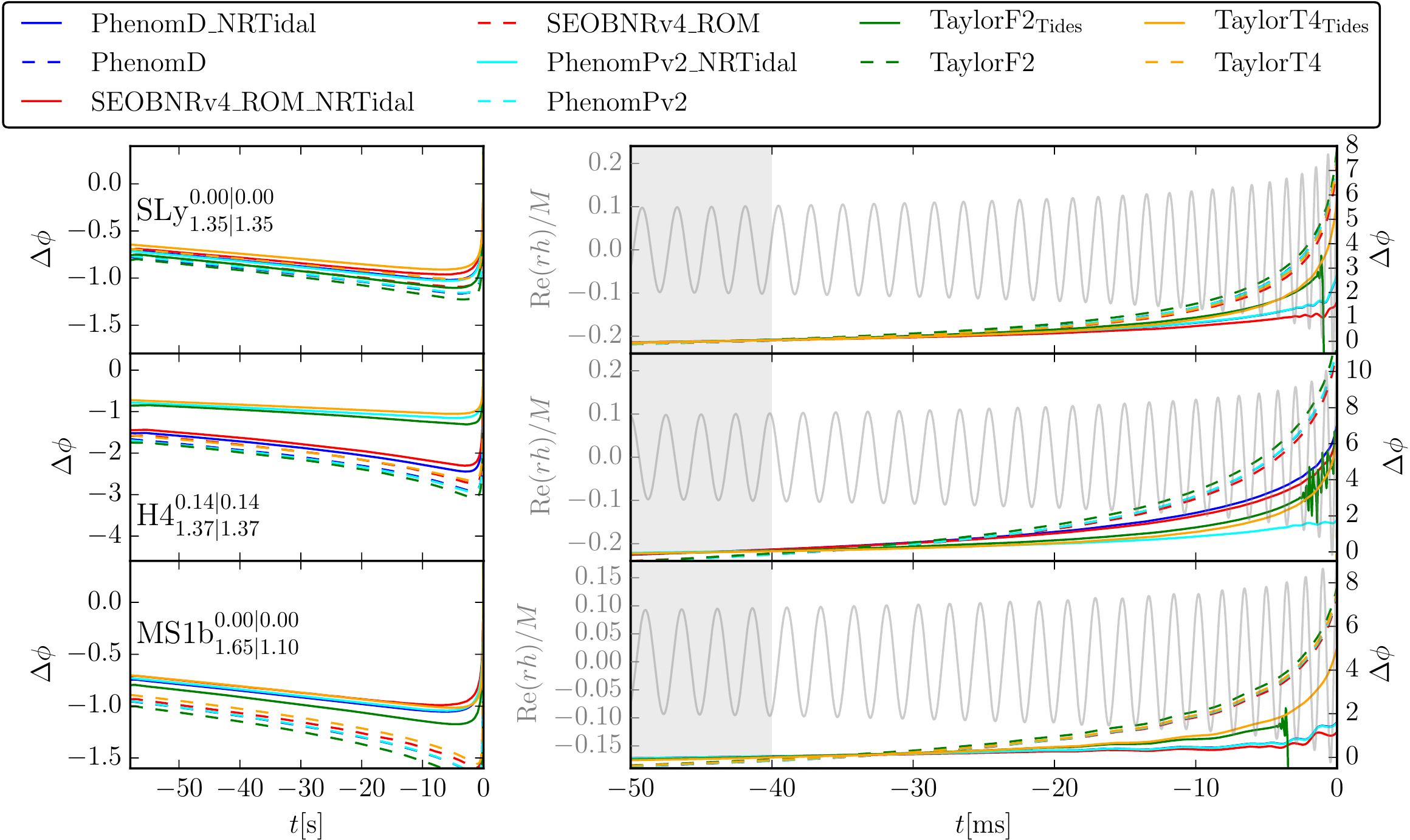}
\caption{
Time domain dephasing $\Delta \phi$ between waveform models and hybrids for
${\rm SLy}_{1.35|1.35}^{0.00|0.00}$ (top panels),
${\rm H4}_{1.37|1.37}^{0.14|0.14}$ (middle panels),
${\rm MS1b}_{1.65|1.10}^{0.00|0.00}$ (bottom panels), similar to Fig.~\ref{fig:time_domain}
but this time aligned in the time interval $t \in [-50~{\rm ms},-40~{\rm ms}]$
before merger, cf.~gray shaded region.}
\label{fig:time_domain_late}
\end{figure*}

We also analyze the performance of the waveform approximants
focusing on the last orbits before merger.
For this purpose we align the waveforms in the time
interval $t \in [-50~{\rm ms},-40~{\rm ms}]$ before the merger.
Most \ac{NR} simulations have an inspiral shorter than $50~{\rm ms}$,
see, e.g., \cite{Hotokezaka:2016bzh,Haas:2016cop,Kiuchi:2017pte,Dietrich:2017xqb} for exceptions.
Consequently, the following analysis is similar to assessing the quality of the waveform approximant
purely based on \ac{NR} simulations.
Let us further emphasize that aligning different waveforms artificially in a regime in which they
disagree can lead to spurious artifacts. For this purpose, we also present the dephasing over the
entire region in the left panels of Fig.~\ref{fig:time_domain_late}.
As done previously, we discuss the three representative examples individually.

\paragraph*{Non-spinning, equal mass} (${\rm SLy}_{1.35|1.35}^{0.00|0.00}$):
Once the waveforms are aligned in the late inspiral, a clear separation
between waveform models that incorporate tidal effects and ones that do not takes place.
All non-tidal approximants have phase differences of the order of $>6~{\rm rad}$ at merger.
The \ac{PN} approximants \TaylorFT and \TaylorTT also have phase differences with respect to the
hybrid of the order of $\sim 5~{\rm rad}$; additionally, \TaylorFT stops before the actual merger.
The \NRtidal models achieve phase accuracies of $\lesssim 2~{\rm rad}$
before the merger, with \SEOBNRROMNRtidal performing best.
In light of the phase differences before the alignment window, we find that
$\Delta \phi$ is negative and of the order of $\sim 1~{\rm rad}$.

\paragraph*{Spinning, equal mass} (${\rm H4}_{1.37|1.37}^{0.14|0.14}$):
We find that for this configuration it is not possible to
align the non-tidal waveforms and the hybrid waveform in a sensible way: in other words,
non-tidal waveforms cannot describe the \ac{BNS} system
at times about $\sim10$ orbits before the merger.
Furthermore, Fig.~\ref{fig:time_domain_late} (middle row)
emphasizes again the importance of the spin-induced and EOS dependent quadrupole
term incorporated in the \PhenomPNRtidal, \TaylorFT, \TaylorTT models.
Overall, for this configuration the \PhenomPNRtidal model performs best with a phase difference
of about $2~{\rm rad}$ at merger.

\paragraph*{Non-spinning, unequal mass} (${\rm MS1b}_{1.65|1.10}^{0.00|0.00}$):
As for the previous case, the non-tidal waveform models do not allow a proper alignment
within the time interval $t \in [-50,-40]~{\rm ms}$.
This is due to the large tidal effects for this particular configuration, which are driven by an effective
tidal coupling constant of $\kappa_{\rm eff}^T=279.4$.
Additionally, we note that the performance of the \TaylorTT model
is worse than the \TaylorFT model for unequal masses and that
the \TaylorFT model stops a few cycles before the actual merger.
The best performances are achieved by the \PhenomPNRtidal, \PhenomDNRtidal and \SEOBNRROMNRtidal
approximants, with a phase difference are merger below $2~{\rm rad}$.

\subsection{Precessing Waveform Comparison}
%==========================
% Fig.8: precessing waeform
%==========================
\begin{figure}[t]
\includegraphics[width=0.495\textwidth]{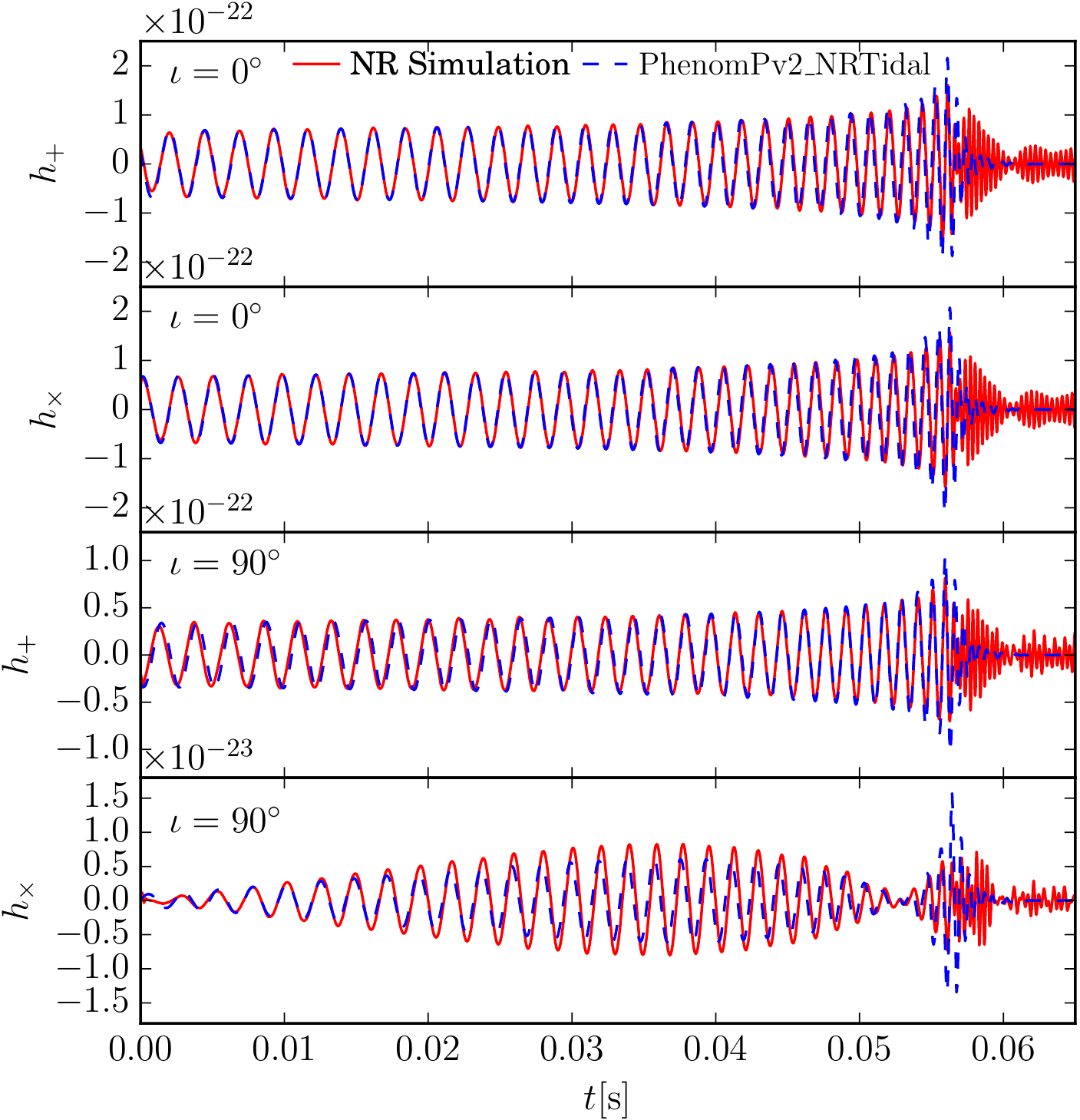}
\caption{
  \ac{GW} strain for a precessing \ac{NR} simulation of~\cite{Dietrich:2017xqb} (red, solid)
  for SLy EOS (see text for details) and the \PhenomPNRtidal model (blue, dashed).
  Results for zero-inclination (face on) are shown in the two top panels;
  results for an inclination of $90^\circ$ (edge on) are shown in the bottom panels.
  We assume a distance to the binary of $100~{\rm Mpc}$.}
\label{fig:TD_precession}
\end{figure}

As a final check, we test the performance of \PhenomPNRtidal
for a precessing \ac{BNS} configuration. Due to the absence of a tidal \ac{EOB} model
including precession effects, we restrict our analysis to the last 15 orbits covered by
the \ac{NR} simulation of~\cite{Dietrich:2017xqb}.
The specific configuration we investigate consists of \acp{NS} with
masses $M_A=1.3553$ and $M_B=1.1072$, tidal deformabilities
$\Lambda_A =382.3$ and $\Lambda_B=1308.6$, dimensionless spins
$\chi_A = (-0.077,-0.077,-0.077)$ and $\chi_B = (-0.089,-0.089,-0.089)$,
and SLy EOS.
The \ac{NR} simulation starts at a \ac{GW} frequency of $392~{\rm Hz}$.

We present the waveform strain for an inclination of $0^\circ$ in the top two panels
of Fig.~\ref{fig:TD_precession}, and for an inclination of $90^\circ$ in the bottom panels.
We assume a distance to the system of $100~{\rm Mpc}$ as in~\cite{Dietrich:2017xqb}.
Overall we find phase differences for $\iota=0^\circ$ below one radian,
which is of the order of the uncertainty in the \ac{NR} simulation~\cite{Dietrich:2017xqb}.
To compare the waveforms, we vary the initial frequency for the \PhenomPNRtidal model
and align the waveforms for $\iota=0^\circ$ at the peak amplitude
followed by an additional time shift to minimize the phase difference.
The same overall time shift is then also employed for the $\iota =90^\circ$ orientation.

Considering the amplitude difference, we do find that the \PhenomPNRtidal waveform
has a larger amplitude close to the merger, which is caused by the missing amplitude
corrections due to tidal effects in the \NRtidal approach.
In the bottom two panels, GW polarizations $h_+,h_\times$ for an inclination of $\iota=90^\circ$ are presented.
The plus-polarization shows a similar trend as for $\iota=0$, i.e,
the amplitude of \PhenomPNRtidal is overestimated, while the overall phasing is in good agreement.
Finally, for $\iota=90^\circ$ precession effects are clearly visible in $h_\times$.
The precession cycle is recovered by the \PhenomPNRtidal waveform which
verifies the assumption that tidal effects and precession effects decouple at LO.
Interestingly, throughout the inspiral the amplitude seems to
be underestimated by the \PhenomPNRtidal model, while the opposite happens at
the moment of merger.
Additionally, we also find that for all inclinations and polarizations
the merger time (conventionally taken as the maximum of the amplitude)
of the \PhenomPNRtidal model determined by Eq.~\eqref{eq:Momega_mrg}
is consistent with the merger of the \ac{NR} simulation.

Overall, bearing in mind the difficulties and uncertainties in the extraction
of spin from the \ac{NR} simulations and the short length of the (single)
\ac{NR} waveform, we conclude that the \PhenomPNRtidal model also seems
able to deliver a consistent representation of the waveform of precessing
systems up to merger.

%==============================================================
\section{Systematics effects in NRTides and their implications}
\label{sec:nrtides_systematics}
%==============================================================

\subsection{Analytical tidal knowledge beyond next-to-leading order
         and its relation with NRTides}
\label{sec:tidesbeyondNLO}

We want to stress that while the \NRtidal approximant is
constructed to reproduce the known NLO tidal knowledge,
analytical knowledge beyond NLO exists~\cite{Damour:2012yf}.
The particular choice for the form of Eq.~\eqref{eq:fitT}
and the restriction to NLO was made because of simplicity and 
to allow a smaller number of parameters in the rational
function used for fitting.

However, analytical information beyond NLO incorporated for example 
in state-of-the-art tidal \ac{EOB} models like
\TEOBResumS have been important to achieve good agreement between \ac{EOB} and \ac{NR} 
waveforms~\cite{Dietrich:2017feu}.
More precisely, the analytical tidal information currently available
and relevant here is: the full next-to-next-to-leading order tidal
contribution to the interbody EOB interaction potential computed in~\cite{Bini:2012gu}
(i.e., formally a 7PN contribution) as well as (ii) tail terms that can be obtained,
at arbitrary PN order, by expanding the resummed tail factor entering
the resummed EOB waveform~\cite{Damour:2007xr,Damour:2012yf}
and (iii) gravitational-self-force contributions to the interaction potential
obtained at high PN order and suitably resummed~\cite{Bini:2014zxa}.
In particular, putting together some of the available analytic information,
Ref.~\cite{Damour:2012yf} obtained the tidal phase at global
7.5PN order that is incorporated in \TaylorF and that is fully known
analytically except for a 7PN waveform amplitude coefficient.
In this respect it was pointed out that such, yet uncalculated,
7PN tidal coefficient $\beta_2^{22}$ entering the quadrupolar waveform
is very likely negligible not only with respect to the other tidal ones
(notably the dynamical ones),
but also with respect to the corresponding 2PN point-mass coefficient.
The  arguments of Ref.~\cite{Damour:2012yf} illustrate that it might be
possible to improve the fitting ans\"atze mentioned above in
Eqs.~\eqref{eq:fitT}-\eqref{eq:Pnrtides} by imposing not only the 6PN term,
but also the 6.5PN and 7.5PN ones (that are analytically fully known) as
well as the 7PN one that is currently lacking the waveform amplitude
contribution $\beta_2^{22}$ mentioned above\footnote{In this respect, one
  has to remind that one could obtain the TaylorF2 tidal approximant
  expanding the EOB analytic phasing to even higher PN order.}.

As an example that illustrates how the current fits,
Eqs.~\eqref{eq:fitT}-\eqref{eq:NRTidal},
differ from the analytically known expression, let us expand
$P_{\rm \Psi_{2.5PN}}^{\rm NRTidal}$ in powers of $x$ up to 2.5PN order.
One finds
\begin{equation}
  \label{eq:nrtides_exp}
P_{\rm \Psi_{2.5PN}}^{\rm NRTidal}=1 + \dfrac{3115}{1248}\,x -
4.22 x^{3/2}+23.32\,x^2 - 111.84 x^{5/2},
\end{equation}
while the analytically expression of Ref.~\cite{Damour:2012yf},
restricted to the equal-mass case, reads
\begin{align}
  \label{eq:psiT_exact}
  \hat{\Psi}^{\rm T}_{\rm 2.5PN}&=1 + \dfrac{3115}{1248}x-\pi x^{3/2} + \left(\dfrac{28024205}{3302208}+\dfrac{20}{351}\beta_2^{22}\right)x^2\nonumber\\
  &-\dfrac{4283}{1092}\pi x^{5/2} \approx 1 + \dfrac{3115}{1248}x-\pi x^{3/2} \nonumber\\
  &+ \left(8.491 + 0.057\beta_2^{22}\right)x^2 - 12.32 x^{5/2}.
\end{align}
One sees here that the (relative) 1.5PN tidal term incorporated
in \NRtidal is about 30\% smaller than the correct analytical one,
while the 2.5PN one is even 9 times smaller. Assuming, as argued in
Ref.~\cite{Damour:2012yf}, that the contribution due to the yet uncalculated
waveform amplitude coefficient $\beta_{22}^2$ can be neglected, the 2PN term
is approximately 2.7 times larger than the corresponding analytical value.
This illustrate the strong ``effectiveness'' of the \NRtidal model already
in the PN-regime, with high-order (effective) PN terms that are required to
fix the imperfect value of the low PN ones. The lack of the correct low-frequency
behavior beyond NLO is per se not a big concern as the approximant should
always be used as a whole; still our analysis illustrates its effective
nature that should be kept in mind.
Consequently, although the current \NRtidal approximant yields rather small
mismatches, for current standards, with \TEOBResumS-based hybrids, we plan
to improve the \NRtidal model in the near future by including beyond-NLO effects.

%============================================
\subsection{Gauge-invariant phasing analysis}
\label{sec:low_freq}
%============================================

\subsubsection{Contributions due to tidal effects}

\begin{figure}[t]
\includegraphics[width=0.495\textwidth]{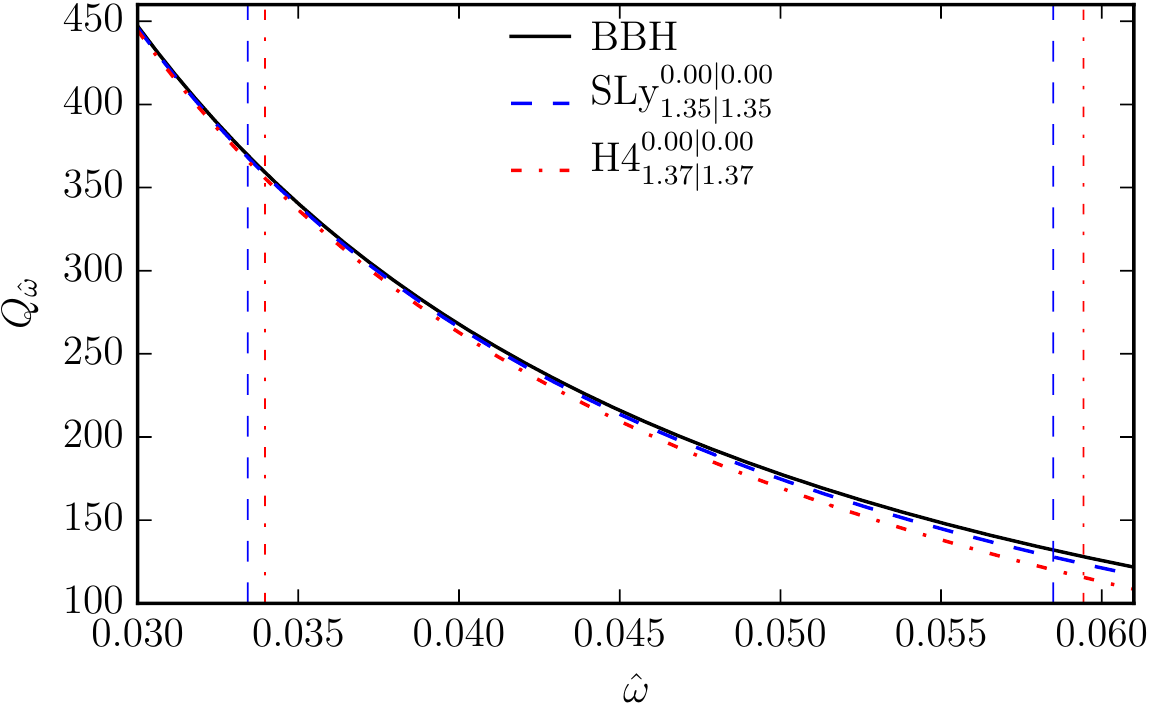}
\caption{Computation of the invariant characterization of the
  phasing of \TEOBResumS, $\Qw$, Eq.~\eqref{Qwdef} for an
  equal-mass, nonspinning, BBH and for two of the
  configurations considered. The effect of tides pushes the BBH
  curve down. Differences that looks small on this scale actually
  correspond to several radians accumulated in phase difference.
  The vertical dashed lines refer to 400 and 700Hz 
  for the two different systems.}
\label{fig:Qomg_basics}
\end{figure}

Let us discuss, from a different perspective, how the tidal phasing yielded
by the \NRtidal model compares with the one of the non-hybridized \TEOBResumS model.
This comparison is especially interesting at low frequencies, a regime that cannot
be touched by NR simulations.
To be conservative, we shall investigate and discuss this comparison up to
dimensionless GW frequency $\hat{\omega}=0.06$, which is the upper limit of the
frequency interval where the hybridization is done. 

To do the comparison in a straightforward way, we build on previous
work~\cite{Baiotti:2010xh,Baiotti:2010xh,Bernuzzi:2012ci} by
using the dimensionless function $Q_\hw$~\cite{Damour:2000gg},
defined as
\begin{equation}
  \label{Qwdef}
 \Qw= \frac{\hw^2}{\partial_t{\hw}}.
\end{equation}
This function has several properties that will be useful in
the present context. First, its inverse can be considered as an adiabatic
parameter $\epsilon_{\rm adiab}=1/Q_{\hw} = \partial_t\hw/\hw^2$ whose
magnitude controls the validity of the stationary phase approximation (SPA)
that is normally used to compute the frequency-domain phasing of PN approximants
during the quasi-adiabatic inspiral. Thus, the magnitude of $\Qw$ itself tells
us to which extent the SPA delivers a reliable approximation to the exact
Fourier transform of the complete inspiral waveform, that also incorporates
nonadiabatic effects. Let us recall~\cite{Damour:2012yf} that, as long as
the SPA holds, the phase of the Fourier transform of the time-domain quadrupolar
waveform
\be
\tilde{h}_{22}(f)\equiv \tilde{A}(f)e^{-{\rm i}\Psi(f)}
\ee
is simply the Legendre transform of the quadrupolar time-domain phase $\phi(t)$,
that is
\be
\Psi_{\rm SPA}(f) = 2\pi f t_f - \phi(t_f) - \dfrac{\pi}{4},
\ee
where $t_f$ is the solution of the equation $\hw(t_f)=2\pi f$.
Differentiating the above equation one then finds
\be
\dfrac{d^2\Psi_{\rm SPA}}{d\hw_f^2}\hw_f^2 =\Qw (\hw_f)
\ee
where now $\hw_f=2\pi f$ is the Fourier domain circular frequency that
coincides, because of the SPA, with the time-domain frequency $\hw(t)$.
Second, the integral of $Q_{\hw}$ per logarithmic frequency yields the phasing
accumulated by the evolution on a given frequency interval $(\hw_L,\hw_R)$, that is
\be
\Delta \phi_(\hw_L,\hw_R)\equiv \int_{\hw_L}^{\hw_R}Q_{\hw} d\log\hw.
\ee
Additionally, since this function is free of the two ``shift ambiguities'' that
affect the GW phase (either in the time or frequency domain), it is perfectly suited
to compare in a simple way different
waveform models~\cite{Baiotti:2010xh,Baiotti:2010xh,Bernuzzi:2012ci,Damour:2012ky,Bernuzzi:2014owa}.
%---------------------------
% Differences between Qomg's
%---------------------------
\begin{figure}[t]
\includegraphics[width=0.495\textwidth]{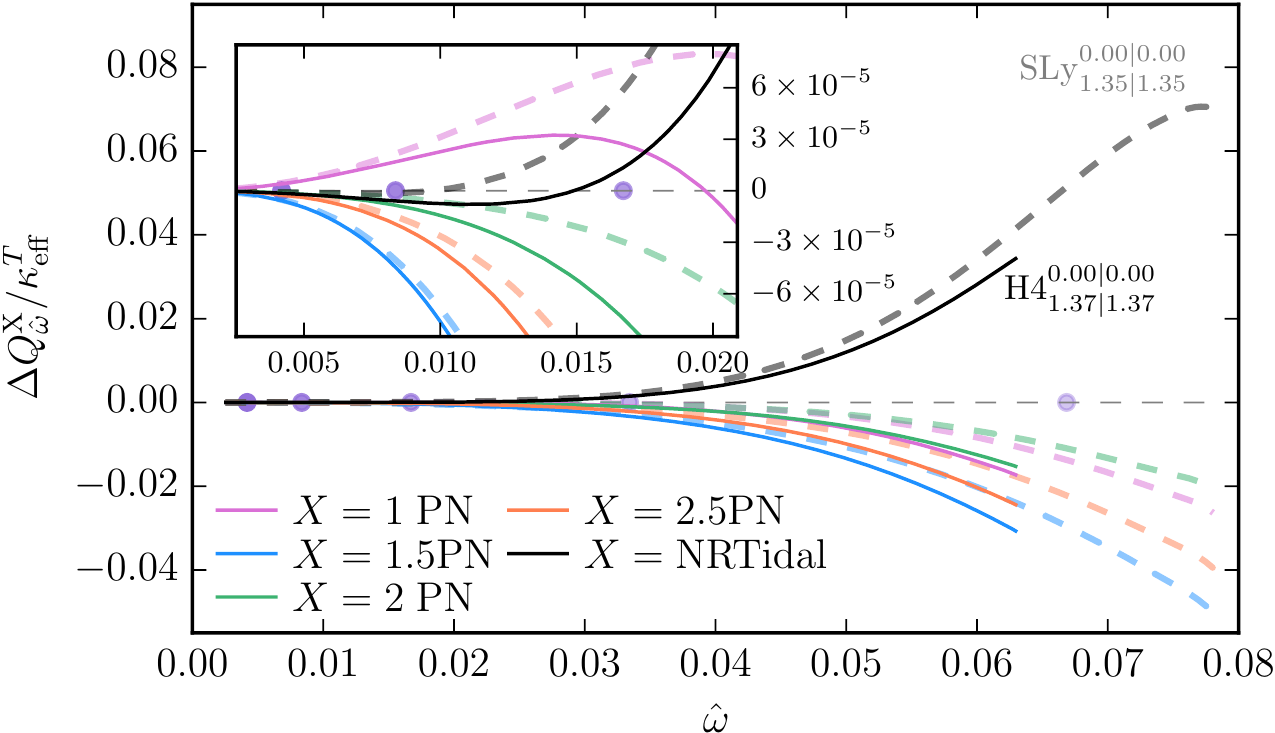}
\caption{ Gauge-invariant description of the tidal phasing for SLy$_{1.35|1.35}^{0.00|0.00}$
  (dashed lines) and H4$_{1.37|1.37}^{0.00|0.00}$ (solid lines). Shown are differences
  between the \TEOBResumS model and different approximants (labeled with $X$) 
  $\Delta Q_{\hat{\omega}}^{\rm X} = Q_{\hat{\omega}}^{\rm TEOBResumS}-Q_{\hat{\omega}}^{\rm X}$
  rescaled by $\kappa_{\rm eff}^T$. The inset shows $\Delta Q_{\hat{\omega}}^{\rm X}$
  for lower frequencies. The purple markers correspond to frequencies
  $(50,100,200,400,800)$~Hz for the SLy$_{1.35|1.35}^{0.00|0.00}$ case.}
\label{fig:TD_early}
\end{figure}
%--------------------------------
We start by computing $Q_\hw$ directly from the time-domain phasing of \TEOBResumS.
We consider two, equal-mass, configurations SLy$_{1.35|1.35}^{0.00|0.00}$ and H4$_{1.37|1.37}^{0.00|0.00}$
starting at 30Hz as well as the corresponding BBH one. Though the calculation of
$Q_\hw$ is, per se, straightforward, since one only has to compute time-derivatives
of $\phi(t)$, in practice there are subtleties that one has to take into account.
First of all, any residual eccentricity related to slightly inconsistent set up
of the initial data of \TEOBResumS will show up as oscillations in the curve
(with typically larger amplitudes the lower the initial frequency is) preventing
one from using this diagnostics for quantitative comparisons. To avoid this, \TEOBResumS
implements post-post-circular initial data~\cite{Damour:2012ky} that are able to
deliver eccentricity-free evolutions. In addition, the time-domain oversampling
of the inspiral may result in high-frequency numerical noise from the computation
of numerical derivatives, that typically hides the low-frequency behavior of the
curve, preventing, for instance, the meaningful computation of differences with
PN approximants in the PN regime ($\simeq 10-30$Hz, see below). To overcome
this difficulty, one has to properly downsample and smooth the raw output
of \TEOBResumS. Figure~\ref{fig:Qomg_basics} shows together the $Q_\hw$
computed for the three equal-mass
binaries, one for BBHs, where tidal effects are set to zero\footnote{This is obtained running \TEOBResumS without
next-to-quasi-circular correction to the waveform and flux~\cite{Nagar:2017jdw},
for consistency with the tidal part where these effects are not included at all.},
(black, solid) and two BNSs, SLy$_{1.35|1.35}^{0.00|0.00}$ (blue, dashed)
and H4$_{1.37|1.37}^{0.00|0.00}$ (red, dash-dotted) up to $\hw\approx 0.06$. The vertical lines
mark 400Hz and 700Hz for SLy$_{1.35|1.35}^{0.00|0.00}$ (blue) and H4$_{1.37|1.37}^{0.00|0.00}$ (red). 
The plot synthetically illustrates two things: (i) the effect of the tidal effects on
phasing, that is attractive, is to push the curve down, 
(ii) the numerical value of the curve up to $\hw\simeq 0.06$ is of order 100.
Since then $\epsilon_{\rm adiab}\sim 0.01$, consistently with~\cite{Damour:2012ky},
we conclude that in this regime the SPA is still a valid approximation and as such
it is meaningful to compare the so-computed time-domain $Q_\hw$ with the corresponding
ones obtained from the frequency domain approximants. Following precisely the same
reasoning of \cite{Damour:2012ky}, we extract  the tidal part of $Q_\hw$ from \TEOBResumS,
to compare it directly with the corresponding ones obtained from \NRtidal,
Eq.~\eqref{eq:NRTidal}, or the PN-expanded one, Eq.~\eqref{eq:psiT_exact}.
We computed the tidal part of the \TEOBResumS $Q_\hw$ as
\be
Q_\hw^{T_{\rm TEOBResumS}}\equiv Q_\hw-Q^0_\hw
\ee
where $Q^0_\hw$ indicates the point-mass curve. We present our results in terms of
differences between $Q_\hw^{\rm T_{TEOBResumS}}$ and \NRtidal, Eq.~\eqref{eq:NRTidal} or
various PN truncations of Eq.~\eqref{eq:psiT_exact}, i.e., we define the quantity
\be
\Delta Q^X_\hw = Q_\hw^{T_{\rm TEOBResumS}}-Q_\hw^{T_X}
\ee
that is shown, rescaled by $\kappa_{\rm eff}^T$, in Fig.~\ref{fig:TD_early} for
SLy$_{1.35|1.35}^{0.00|0.00}$ (dashed lines) and H4$_{1.37|1.37}^{0.00|0.00}$ (solid lines).
On the same plot, the purple markers correspond to frequencies (50,100,200,400,800)~Hz
for SLy$_{1.35|1.35}^{0.00|0.00}$.
The plot illustrates the following facts.
\begin{itemize}
\item[(i)] In the ``early'' frequency range $f\lesssim 150~{\rm Hz}$ (see inset), for both
configurations the difference between  the \NRtidal approximant and the \TEOBResumS model
is below $10^{-5}$ and always smaller than for the PN approximants. This keeps being small
also after multiplication by $k^T_{\rm eff}\approx10^2$, which assures that the two models just
negligibly dephase up to 150~Hz. This confirms the quality of the calibration  
of the \NRtidal model to \TEOBResumS, illustrating that the fit of the high-frequency
part, probably thank to having imposed the correct LO tidal behavior, did not lead to
dramatic uncertainties at low frequencies.
The same plot also shows that even at frequencies $\sim 50~{\rm Hz}$, i.e., several
hundred orbits before the actual merger there are noticeable differences between
the tidal PN approximants and \TEOBResumS. This emphasizes that  the PN regime is
not yet met there (and it in fact extends below 30~Hz) so that one should not,
in principle, restrict to the use of simple PN-expanded descriptions of tidal
effects. One sees in this respect that the 1PN tidal approximant overestimates
the effects (i.e., the inspiral is accelerated), while all the other
approximant underestimate them with respect to \TEOBResumS.

\item [(ii)] when moving to higher frequencies $f\gtrsim 150$~Hz, it is
found that \NRtidal yields stronger tidal effects with respect to the
\TEOBResumS baseline. Once multiplied by $\kappa^T_{\rm eff}$, the differences shown
in the figure become $\sim 2$ for SLy$_{1.35|1.35}^{0.00|0.00}$ and order $\sim 6$
for  H4$_{1.37|1.37}^{0.00|0.00}$ at 700Hz, which implies an accumulated phase
difference up to that frequency of the order of a radian. 
By contrast, several studies~\cite{Bernuzzi:2012ci,Bernuzzi:2014owa,Dietrich:2017feu,Nagar:TEOBResumS}
have illustrated the high degree of compatibility between state-of-the-art
NR simulations and \TEOBResumS (or analogous, EOB-based, waveform
models~\cite{Hinderer:2016eia,Steinhoff:2016rfi})
up to, and often beyond, $\hw=0.06$. As a consequence when \NRtidal is used to
extract the tidal parameters from actual GW signals, one may 
expect to get smaller tidal deformabilities with respect to those
predicted by PN or EOB approximants in order to compensate for the
aforementioned systematics in the post-LO tidal correction yielded by
$P_\Psi^{\rm NRTidal}$, Eq.~\eqref{eq:NRTidal}. This calls for further
adjustment to the \NRtidal model in the range $150-800$~Hz in order
to  increase its capability to catch tidal effects in
throughout the entire inspiral.

\item[(iii)] The third piece of information yielded by Fig.~\eqref{fig:TD_early}
  is that the dashed and solid lines do not coincide. This is not surprising,
  and actually expected, since $\kappa_2^T$ (or $\kappa^T_{\rm eff}$) only takes into
  account leading-order (conservative) tidal effects, while their incorporation in
  a state-of-the-art EOB model is more 
  complicated~\cite{Damour:2009wj,Vines:2011ud,Bini:2012gu,Damour:2012yf,Steinhoff:2016rfi}.
  For example, the presence of $\ell=3$ and $\ell=4$ tidal corrections (that become
  more important the more deformable the star is) as well as of additional, mass-ratio-dependent,
  effects~\cite{Damour:2009wj,Bini:2012gu} that were not incorporated in the simplified (mass-ratio independent)
  fit given by $P_\Psi^{\rm NRTidal}$ in Eq.~\eqref{eq:NRTidal}. Still, as it should, the linearity
  in the tidal parameter is recovered correctly in the medium-low frequency regime 30-100Hz.
  
\end{itemize}

\subsubsection{Contributions due to the spin-induced quadrupole moment}
\label{sec:checkSS}

\begin{figure}[t]
  \includegraphics[width=0.495\textwidth]{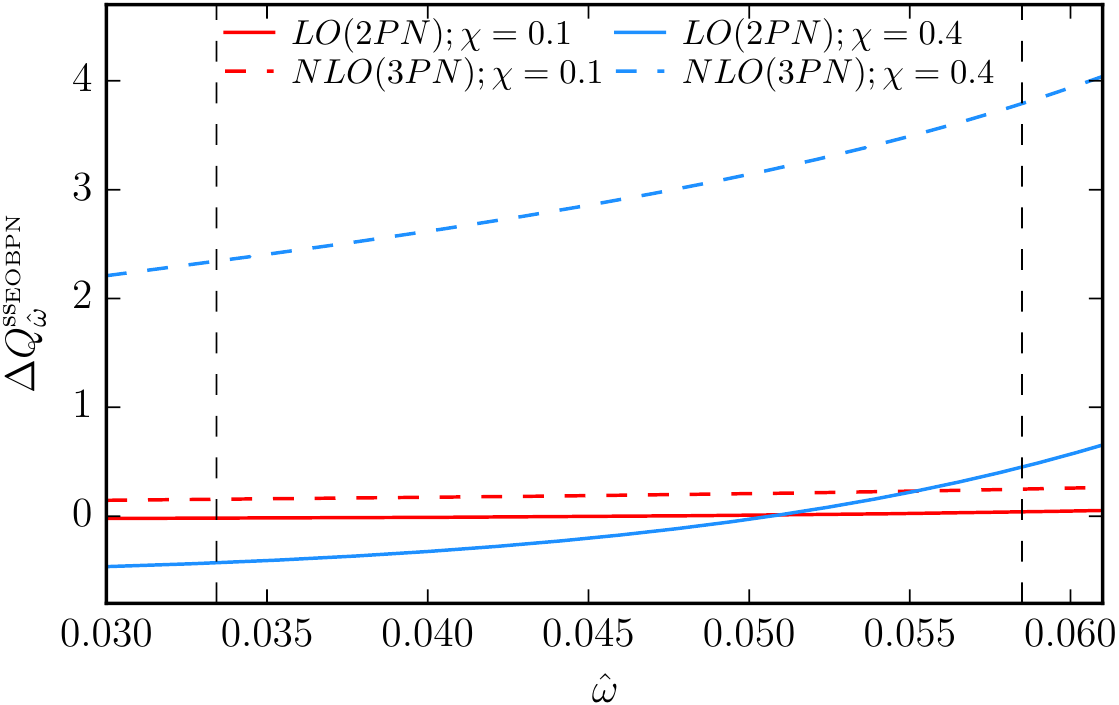}
  \caption{Gauge-invariant description of the self-spin contribution to the phasing
    for an equal-mass binary with $M^A=M^B=1.35$ employing the MS1b EOS.
    The spin magnitudes are $\chi^A=\chi^B=0.1$ (red) and $\chi^A=\chi^B=0.4$ (blue).
    The vertical dashed lines refer to 400 and 700Hz. The plot shows the difference
    $\Delta Q^{\rm ss_{EOBPN}}_{\hat{\omega}}$ between the self-spin $Q_\hw^{\rm ss_{EOB}}$ obtained with
    the \TEOBResumS model and the corresponding PN-expanded expressions at LO
    (2PN, solid lines~\cite{PhysRevD.57.5287}) or NLO
    (3PN, dashed lines\cite{Bohe:2015ana}), as actually implemented in \PhenomPNRtidal.
    For astrophysically motivated values of the spins, $0.1$, the effect of the PN
    NLO term is still compatible with \TEOBResumS, that is constructed using only
    LO self-spin information in the Hamiltonian and in the flux. By contrast, it is not
    the case for larger values of the spins, so that a full assessment of the approximant
    in that regime calls for modification of the self-spin content of \TEOBResumS.}
\label{fig:Qomg_spin}
\end{figure}

Similarly to the discussion about the tidal contribution to the GW phasing, 
we also want to discuss the imprint of the quadrupole-dependent spin-spin 
effects as they are implemented in \PhenomPNRtidal. As mentioned above,
\PhenomPNRtidal, as well as \TaylorFT, implement EOS-dependent self-spin
terms up to NLO, i.e. that include both the LO, 2PN-accurate,
term~\cite{PhysRevD.57.5287} and as well as the 3PN-accurate term that
can be deduced from Ref.~\cite{Bohe:2015ana}. By contrast, \TEOBResumS
only implements self-spin information at LO in both the Hamiltonian and
the flux\footnote{We recall that, since the EOB waveform is resummed,
  other terms, notably the tail ones, are effectively included in
  the \TEOBResumS expression for the flux, while they are not part of the PN approximants.}.
Since we used the \TEOBResumS model as baseline to validate the performance
of \PhenomPNRtidal, we need to check that the effect of the NLO terms present
in \PhenomPNRtidal is not dramatically stronger than the LO ones over the
explored parameter range.
To do so, we isolate the self-spin contribution, $Q_\hw^{\rm ss}$, to the phasing
of \TEOBResumS and compare it with the LO and NLO contributions in \PhenomPNRtidal.
We follow the procedure discussed in Ref.~\cite{Nagar:TEOBResumS},
where $Q_\hw^{\rm ss}$ is extracted by subtracting from the total $Q_\hw$ the corresponding
one obtained putting the quadrupolar deformation parameters, $C_Q$, to zero.
In order to get an upper limit to the effect, we chose and equal-mass configuration
$M^A=M^B=1.35M_\odot$ for the MS1b EOS, due to the large (unphysical) deformability,
and we put either $\chi^A=\chi^B=0.1$ or $\chi^A=\chi^B=0.4$. The value of the 
quadrupolar parameter induced by rotation in this case is $C_{QA}=C_{QB}\simeq 8.396$.
For the sake of comparison, we recall that $C_{QA}=C_{QB}=1$ for the BH case,
and $C_{QA}=C_{QB}\simeq 5.49$ for a SLy configurations with the same individual masses.
Figure~\ref{fig:Qomg_spin} depicts the difference $\Delta Q_\hw^{\rm ss}=Q_{\hw}^{\rm ss_{\rm EOB}}-Q_\hw^{\rm ss_{\rm PN}}$
for both the LO (solid) and NLO (dashed) PN truncations and
for the two values of the spins. The two vertical lines refer to 400 and 700Hz.
The figure illustrates several facts. Focusing first on the, astrophysical motivated,
small spin case, one sees that both the LO and NLO truncations of the PN approximants
are essentially consistent with \TEOBResumS. More precisely, the ${\rm EOB-PN}$
phase difference accumulated between 20~Hz ($\hw\simeq 0.00251$, not represented in the figure)
and $\hw =0.06$ is $\Delta\phi^{\rm ss}_{\rm LO}=-0.063$ in the LO case
and  $\Delta\phi^{\rm ss}_{\rm NLO}=+0.348$ in the NLO case. At a more detailed
level~\cite{Nagar:TEOBResumS}, one finds that the NLO term used in the \PhenomPNRtidal
model overall increases the effect of the self-spin contribution such that it is globally
stronger, and thus more attractive, than \TEOBResumS. When the spin is as large
as $0.4$, one finds the same qualitative behavior, but the accumulated dephasings
become unacceptably large, with $\Delta\phi^{\rm ss}_{\rm LO}=-1.29$ and $\Delta\phi^{\rm ss}_{\rm NLO}=+5.31$.
Although we are prone to think that such accumulated phase differences are mostly
due to the non-robust behavior of the NLO PN approximant, we are currently not able to
prove this, as to do so would require to consistently incorporate the EOS-dependent
NLO information both in the conservative and nonconservative sector of the model.
On a positive side, however, one has to remember that the results of Fig.~\ref{fig:Qomg_spin}
refer to an EOS that has very reduced probability of exist in Nature, which
seems to select softer EOS models with smaller rotation-induced quadrupole moment~\cite{TheLIGOScientific:2017qsa}.
This suggests that the effect of the NLO terms in \PhenomPNRtidal is subdominant with
respect to the LO ones and the comparisons with the \TEOBResumS baseline should be
considered reliable.

\section{Summary and future perspectives}
\label{sec:conclusion}

We described in detail the implementation of the \NRtidal models in
the LSC Algorithm Library Suite, along with detailed tests of these
new waveform approximants. Our study is timely, as such approximants
were already employed to estimate the properties of the source of
the \ac{GW} signal GW170817 and they will be employed for further
analysis of the system.
To validate the performance of the \NRtidal models, we computed mismatches,
frequency-domain phase differences, and time-domain phase differences
between the different waveform models and target BNS waveforms constructed
by hybridizing tidally improved EOB waveforms, obtained via the
\TEOBResumS model~\cite{Nagar:TEOBResumS}, with high-resolution \ac{NR}
simulations covering the last orbits of the inspiral up to merger.
The additional new theoretical input incorporated in \TEOBResumS is that
the model blends together, in a resummed fashion, tidal and spin effects,
notably including EOS-dependent self-spin effects.
This special feature not only allows one to asses the performance of \NRtidal
models, but also that of the PN-based description of the self-spin effects
that is present in some of the LAL approximants.

Our main observations are:
\begin{enumerate}[(i)]
\item For spinning BNSs or stiff EOSs, non-tidal approximants fail
  to describe the evolution of the binary system.
  Consequently, every analysis discussing properties of BNS
  systems has to be based on waveform approximants incorporating
  tidal effects.
\item For spinning systems, the inclusion of the spin-induced,
  and EOS-dependent 2PN term~\cite{PhysRevD.57.5287} (and 3PN~\cite{Bohe:2015ana})
  terms in the waveform  approximants proves crucial to reduce
  the mismatches with the \TEOBResumS-driven inspiral waveform when the spin
  magnitudes are $\gtrsim 0.1$~\cite{Harry:2018hke}.
\item For configurations with large unequal masses ($q=1.5$) and/or
  large tidal effects, the performance of the \NRtidal models
  is better than the performance of the \ac{PN} based waveform
  approximants by more than an order of magnitude in terms of mismatches.
\item We compared and contrasted in detail the \NRtidal representation
  of the tidal interaction with the one incorporated in \TEOBResumS.
  We concluded that the \NRtidal model systematically overestimates
  the tidal interaction with respect to \TEOBResumS also in the intermediate
  frequency 150-800~Hz where \TEOBResumS is expected to be fully reliable.
  If, on the one hand, this calls for improvements in the construction
  of \NRtidal, on the other hand, despite the very small mismatches found
  ($<10^{-2}$ of $\sim 10^{-3}$), one should be careful about possible
  systematics in parameter estimation studies brought by the use of \NRtidal.
  Such uncertainties should be properly assessed by means of injection studies
  employing full parameter estimation pipelines.
\item
  Finally, by comparing the \PhenomPNRtidal model also to a
  precessing \ac{NR} simulation we conclude that the model looks
  sufficiently mature to deliver a qualitatively and semi-quantitatively
  consistent representation of the last orbits of precessing BNS systems.
\end{enumerate}

Based on the chosen target waveform set, we find that the
\PhenomPNRtidal model gives the smallest mismatches and phase
differences in the frequency and time domain.
It is currently the only frequency model implemented in
the LALSuite which incorporates phase corrections
based on \ac{NR} and \ac{EOB} tuned tidal effects,
the spin induced EOS dependent quadrupole moment,
and precession of the orbital angular momentum for non-aligned spins.

Although the current implementation of the \NRtidal waveform
approximant is a step towards an efficient modeling of
tidal effects in BNS systems like GW170817, there are
immediate tests and improvements that we want to outline
here to steer future developments of the model. 
The three obvious checks are 
(i) tests of the performance of the \PhenomPNRtidal model
for precessing systems for the entire inspiral,
(ii) tests against different hybrid models based on
other waveform models and \ac{NR} simulations,
and (iii) tests the effect of the waveform
approximant in the context of parameter estimation studies.
Considering (ii) and (iii), we remark that there are ongoing
injection studies to assess systematic uncertainties of waveform
approximants for parameter estimation purposes.
Furthermore, the implementation of the tidal
\ac{EOB} model \texttt{SEOBNRv4T}~\cite{Hinderer:2016eia},
which includes the EOS dependent quadrupole effect,
was recently completed.  This will allow to construct
additional hybrids using models that rely on different
inspiral waveforms. However, there is currently no
possibility to compare the \PhenomPNRtidal model
against precessing systems throughout the entire frequency band
accessible to advanced \ac{GW} detectors.
While the recent progress in \ac{NR} allowed us to validate the performance
of \PhenomPNRtidal during the last $15$ orbits before merger,
there is no tidal \ac{EOB} model which incorporates precession effects
to enable a study of the early inspiral.

Considering possible improvements of the \NRtidal approximants, 
we plan in the near future to (i) incorporate analytical tidal corrections 
to the amplitude of the GW, (ii) include the EOS dependence of
the spin-induced  quadrupole momentum in the \PhenomDNRtidal
and \SEOBNRROMNRtidal waveform models, and most notably,
(iii) try to incorporate analytical knowledge beyond next-to-leading order 
tidal contributions to the \NRtidal approximant to further
improve the performance of the model throughout the entire inspiral. 

\begin{acknowledgments}
  We want to thank Nathan Johnson-McDaniel for his participation in the LIGO review process of the
  \NRtidal models, in particular for his checks of the merger frequency and comparisons
  to BBH systems and NR waveforms.
  We also thank Tanja Hinderer for her work on the review of the \PhenomPNRtidal model
  and Francesco Messina for carefully cross-checking the implementation of the self-spin
  terms in the \TaylorFT model.
  We also thank Sergei Ossokine for helpful discussions about precessing
  compact binary systems.
  We thank Alessandra Buonanno, Nathan Johnson-McDaniel, and 
  Noah Sennett for comments on the manuscript. \\
  TD acknowledges support by the European Union's Horizon
  2020 research and innovation program under grant
  agreement No 749145, BNSmergers.
  SB acknowledges support by the EU H2020 under ERC Starting Grant,
  no.~BinGraSp-714626.
  CM acknowledges support by the European Union's Horizon
  2020 research and innovation program under grant
  agreement No 753115, ACFD.
  KWT, AS, TD and CVDB are supported by the research programme 
  of the Netherlands Organisation for Scientific Research (NWO).
  SK, FO and YS are supported by the Max Planck Society's Independent Research
  Group Grant.
  MH acknowledges support by the Swiss National Science Foundation (SNSF).
  FP acknowledges support by Science and Technology Facilities Council (STFC)
  grant ST/L000962/1 and European Research Council Consolidator Grant 647839.
  RD has been supported by the DFG Research Training Group 1523/2 "Quantum and Gravitational Fields" .

  Computations of the numerical relativity waveforms have been performed performed
  on the supercomputer SuperMUC at the LRZ
  (Munich) under the project number pr48pu,
  the compute cluster Minerva of the Max-Planck
  Institute for Gravitational Physics, and on the Hydra and Draco clusters of
  the Max Planck Computing and Data Facility.
\end{acknowledgments}
\appendix

\bibliography{paper20180406}

\end{document}